\documentclass[aps,prd,reprint,superscriptaddress,nofootinbib,amsmath,amssymb,floatfix]{revtex4-1}

\usepackage{graphicx}
\usepackage{bm}
\usepackage{booktabs}
\usepackage{xcolor}
\usepackage{xspace}
\usepackage[hidelinks]{hyperref}
\usepackage{float}
\hypersetup{colorlinks=true,citecolor=teal,linkcolor=teal,urlcolor=teal}
\newcommand{\adonis}{\textsc{ADoNIS}\xspace}

\graphicspath{{figures/}}

\definecolor{acronym_color}{RGB}{195,105,105}

\begin{document}

\title{
\textcolor{acronym_color}{ADoNIS}:
\textcolor{acronym_color}{A}
\textcolor{acronym_color}{D}ifferentiable generat\textcolor{acronym_color}{O}r of
\textcolor{acronym_color}{N}eutrino
\textcolor{acronym_color}{I}nteraction
\textcolor{acronym_color}{S}amples
}

\author{C\'esar Jes\'us-Valls}
\email{cesar.jesus@cern.ch}
\affiliation{European Organization for Nuclear Research (CERN), 1211 Geneva 23, Switzerland}

\begin{abstract}
\noindent Neutrino interaction generators are central to precision oscillation
analyses, but conventional implementations do not directly provide derivatives
of their predictions with respect to physics parameters, leaving analyses to
obtain this dependence through reweighting, finite differences or external
response models. We present \adonis, a fully differentiable neutrino interaction
event generator that makes event reweighting and its gradients directly
available, propagating exact derivatives through the nuclear ground state,
hard-scattering amplitudes and stochastic intranuclear cascade. Following the
physics choices of ACHILLES, we demonstrate agreement with its predictions
across neutrino, electron and hadron probes, showing that differentiability is
achieved without loss of physical fidelity. We then show how the resulting
Jacobian exposes which measurements and regions of phase space constrain each
parameter and how different probes break degeneracies, providing a quantitative
tool for fit design, model tuning, measurement design and experiment design.
The same differentiable predictions can be used directly in frequentist and
Bayesian inference and carried through detector response to reconstructed
observables for unfolding, while also yielding computational gains. In sum, \adonis provides a practical recipe for constructing differentiable
neutrino event generators, applicable beyond the particular physics models
implemented here, and this paper demonstrates their use across a wide range of tasks
relevant to both theory and experiment.
\end{abstract}

\maketitle


\section{Introduction}

Neutrino oscillation experiments have entered a precision era. The determination of the leptonic charge-parity (CP)-violating phase, the resolution of the
neutrino mass ordering, and the increasingly stringent tests of the
three-flavour paradigm all rest on comparing measured event rates against
predictions at the percent level~\cite{T2K:2025wet, NOvA:2025tmb,
DUNE:2020jqi, Hyper-Kamiokande:2025fci}. Because accelerator neutrino beams
are broad-band, the energy of each interacting neutrino is not known a priori
and must be inferred from the observed final state. That inference is made
through a model of the neutrino--nucleus interaction, so the accuracy of the
model propagates directly into the accuracy of the oscillation measurement.
Interaction modelling is consequently among the leading systematic
uncertainties of the field, and one of the most difficult to control, since its
sources lie in details of nuclear physics that are not directly
observable, and can mimic the very shifts in the oscillation parameters that
are being measured~\cite{Coyle:2025xjk, Liu:2025hpl, Dolan:2026nlr}.

The difficulty is that at $\mathcal{O}(\mathrm{GeV})$ energies, typical of
accelerator neutrino oscillation experiments, several neutrino--nucleus
interaction channels contribute simultaneously and involve a broad range of
nuclear and particle physics. Quasielastic scattering, multinucleon processes and resonance production
depend on the nuclear environment and nucleon structure, while at higher
energies interactions increasingly probe the internal structure of the
nucleon. A realistic prediction must therefore combine
the initial nuclear state, the relevant interaction mechanisms, the transport
of produced hadrons through the nucleus, and the de-excitation of the remnant,
while providing a consistent description across kinematic regimes and target
nuclei.

Every one of these stages is a research front in its own right, pursued by
different groups with different formalisms, approximations, and domains of
validity. Descriptions of the nuclear ground state continue to be
refined~\cite{Wiringa:2013ala,
Lovato:2015qka, Rocco:2018vbf, Weiss:2018tbu, Cruz-Torres:2019fum,
Rocco:2020jlx, Sobczyk:2022ezo, JeffersonLabHallA:2022cit, Sobczyk:2023mey,
Nikolakopoulos:2023zse, Ankowski:2024ntv}. At the primary interaction,
quasielastic and resonant descriptions~\cite{Gonzalez-Jimenez:2014eqa,
Megias:2016lke, Gonzalez-Jimenez:2019ejf, Butkevich:2021sfn,
Franco-Patino:2022tvv, Nikolakopoulos:2023pdw, Franco-Patino:2023msk,
Gonzalez-Rosa:2024udj, McKean:2025khb, Nakamura:2015rta, Nakamura:2016cnn,
Sobczyk:2018ghy, Nikolakopoulos:2022tut, Kabirnezhad:2024cor, Yan:2024kkg,
Gonzalez-Rosa:2023aim, Gonzalez-Rosa:2026cbz}, two-body
currents~\cite{Martini:2009uj, Nieves:2011pp, Bodek:2011ps, Nieves:2012yz,
VanCuyck:2016fab, Gallmeister:2016dnq, Dolan:2019bxf, Sobczyk:2020dkn,
Casale:2025wsg, Casale:2025zac, Martinez-Consentino:2025ebw,
Kasturi:2026oyt, Franco-Munoz:2026hxi}, nucleon form
factors~\cite{Bodek:2007vi, Meyer:2016oeg, RQCD:2019jai,
Alexandrou:2020okk, Meyer:2022mix, Djukanovic:2022wru, MINERvA:2023avz,
Tomalak:2023pdi, Jang:2023zts, MINERvA:2025ygc, Meyer:2026kdl}, coherent
pion production~\cite{Alvarez-Ruso:2007rcg, Alvarez-Ruso:2007kwp,
Amaro:2009vu, Nakamura:2009iq, Saraswat:2016kln, Sogarwal:2022xle}, and
shallow- and deep-inelastic scattering~\cite{NuSTEC:2019lqd,
SajjadAthar:2020nvy, SajjadAthar:2025fhk, Jeong:2023hwe, Ansari:2020xne,
Ansari:2021cao, Zaidi:2021iam, AtharSajjad:2022ipr, Zaidi:2026rxo,
Muzakka:2022wey, Helenius:2021tof, Klasen:2023uqj, Yang:2009zx,
FerrarioRavasio:2024kem} are all subjects of active theoretical and
phenomenological development. The subsequent transport of hadrons through the
nucleus~\cite{Buss:2011mx, Boudard:2012wc, Golan:2012wx, Niewczas:2019fro,
Nikolakopoulos:2022qkq, Ershova:2022jah, Isaacson:2022cwh,
Nikolakopoulos:2024mjj, Butkevich:2025fsf, Yan:2025aau, Isaacson:2025cnk}
and the de-excitation of the residual system~\cite{Ershova:2023dbv,
Abe:2023iwq} are likewise active areas of development.

Making these ingredients consistent across interaction channels and target
nuclei remains an open and actively contested
problem~\cite{NuSTEC:2020nsl, Alvarez-Ruso:2025oak, Dolan:2026dfu,
Dolan:2026nlr}. Event generators are where these ingredients are assembled
into complete final-state predictions for experimental analyses. The field
therefore uses several generators, including
GENIE~\cite{GENIE:2021npt, Andreopoulos:2009rq},
NEUT~\cite{Hayato:2021heg}, NuWro~\cite{Golan:2012wx},
GiBUU~\cite{Buss:2011mx}, and ACHILLES~\cite{Isaacson:2020wlx,
Isaacson:2022cwh}, reflecting different scopes, use cases, and choices of
physics implementation.

Once these ingredients are assembled in an event generator, a separate
challenge remains: using its output to infer physics from data. An event
generator maps a set of physics parameters to simulated final states, which an
analysis then turns into observable predictions that can be compared with
measurements. At truth level, these predictions can be compared with
cross-section measurements to validate interaction models, tune their
parameters and derive constraints. In an experimental analysis, instead, the
simulated events are propagated through the detector and compared with data in
reconstructed kinematics to infer the underlying physics. In either case,
inference requires knowing how the observable predictions change when the
physics parameters are varied. Conventional generators do not provide this
information directly as derivatives, so analyses must obtain it through
additional parameter-response machinery~\cite{Stowell:2016jfr,
Chakrani:2022tey, GENIE:2024ufm, Abe:2025yaa, T2K:2026cqe}. Building and
maintaining this machinery, sometimes tied to particular analysis choices,
slows the adoption of new theoretical developments in experiments and, in
turn, delays the experimental feedback that helps theorists steer further
improvements. That delay matters especially now, when many parts of
neutrino-interaction modelling are being improved at the same time and
experiments increasingly need those developments to keep pace with their
precision.

For parameter inference, one way to work around this limitation is
simulation-based inference, which trains a machine-learning model on repeated
simulations to learn an inverse mapping from the resulting observables back to
the physics parameters~\cite{Cranmer:2019eaq,
Brehmer:2019xox}, an approach recently applied to neutrino-interaction model
tuning~\cite{Tame-Narvaez:2025pwg,Tame-Narvaez:2026zmc}. The way the
predictions change when the physics parameters are varied is, however, already
encoded in the physics calculation itself. A differentiable generator exposes
this response directly, event by event, through exact derivatives. Rather than
learning an approximate inverse relation from repeated simulations, it
provides the derivatives of the forward calculation itself. These derivatives
can then be propagated to arbitrary observables to show which parameters
affect them, where different parameters produce similar effects, and which
datasets add new information, while also supporting parameter inference,
uncertainty estimation and measurement design.

Differentiability concerns how a calculation is implemented rather than which
physics model it represents, and therefore need not favour one model over
another or come at the expense of physical fidelity or interpretability. The
construction presented here is therefore general: it provides a recipe for
making event generators differentiable that is not tied to the particular
interaction models implemented in \adonis. This motivates treating
differentiability as a general capability of future neutrino interaction
generators. To demonstrate this in a realistic setting, this paper presents
\adonis, a fully differentiable neutrino event generator. We use it to show
how differentiability can be maintained throughout the neutrino-interaction
calculation, including through the stochastic intranuclear cascade, and explore
what the resulting gradient information makes possible.

The remainder of this paper is organised as follows.
Section~\ref{sec:adonis} describes the construction of \adonis and how exact
derivatives are propagated from the interaction-model parameters to individual
events and arbitrary observables. Section~\ref{sec:validation} validates the
differentiable implementation against ACHILLES across neutrino, electron and
hadron probes. Section~\ref{sec:jacobian} uses the resulting Jacobian to show
where model parameters affect measured distributions, which combinations are
constrained or degenerate, and how different measurements provide
complementary information. Section~\ref{sec:fitting} demonstrates how the exact gradient information can
be used directly for parameter inference and uncertainty quantification,
including the computational advantages it provides. Finally, Sec.~\ref{sec:unfolding} demonstrates unfolding within the same
differentiable framework, carrying the prediction through detector response
and extracting a cross section from reconstructed data.

\subsection{Differentiable programming in particle physics}
\label{sec:intro_diffprog}

Differentiable programming makes a simulation differentiable in its own
inputs, so that derivatives of the prediction are produced by the program
rather than reconstructed outside it. Automatic
differentiation~\cite{Baydin:2015tfa} decomposes a
program into elementary operations and applies the chain rule through each of
them, giving the exact derivative of the output with respect to any input; in
reverse mode a full gradient costs a small constant multiple of one forward
evaluation, independent of the number of inputs, and higher derivatives follow
by nesting the same construction.

Physics is adopting this formalism
broadly~\cite{Schoenholz:2019jaxmd, Kasim:2021dqc, Weymouth:2023waterlily,
Meunier:2025veros, Liao:2019tensor, Kaiser:2024cheetah, Agol:2021nbody,
Li:2022pmwd}. Within high-energy physics, applications are emerging rapidly.
The stochastic sampling long presumed to obstruct them has proved
tractable~\cite{Arya:2022stochad}, and gradient information has been put to
use across the full range of common experimental tasks: detector
design~\cite{MODE:2022znx}, physics modelling~\cite{Heinrich:2022xfa,
Cranmer:2021gdt, Nachman:2022jbj, BarhamAlzas:2024ggt}, and data
analysis~\cite{Simpson:2022suz}.
In neutrino physics, differentiable frameworks now exist for the two most
widely used detector technologies, liquid argon time projection
chambers~\cite{Gasiorowski:2023tqf} and optical Cherenkov and scintillation
detectors~\cite{Alterkait:2026ocv, CCM:2025dbq}, with demonstrated applications to detector calibration and event reconstruction. \adonis extends this emerging paradigm to the modelling of neutrino interactions.


\section{A Differentiable Neutrino Event Generator}
\label{sec:adonis}

We developed \adonis, \textbf{A} \textbf{D}ifferentiable
generat\textbf{O}r of \textbf{N}eutrino \textbf{I}nteraction \textbf{S}amples,
a JAX-based open-source neutrino event
generator\footnote{\href{https://github.com/cesarjesusvalls/ADoNIS/}{https://github.com/cesarjesusvalls/ADoNIS/}}.
\adonis follows the interaction physics choices of the ACHILLES
generator~\cite{Isaacson:2020wlx, Isaacson:2022cwh, Isaacson:2025cnk},
implemented in a fully differentiable manner, and so provides differentiable
event weights for fully exclusive neutrino--nucleus final states.

ACHILLES is a modern,
open-source\footnote{\href{https://github.com/AchillesGen/Achilles}{https://github.com/AchillesGen/Achilles}},
self-consistent generator built around fully exclusive final states. Its
vertex implements two channels. Quasi-elastic scattering is computed from the
one-body electroweak current in the extended factorisation
scheme~\cite{Isaacson:2022cwh}, with Kelly vector and $z$-expansion axial
nucleon form factors~\cite{Kelly:2004hm, Meyer:2016oeg}. Resonant single-pion production is taken from the ANL--Osaka dynamical
coupled-channels model~\cite{Nakamura:2015rta}, whose meson--baryon amplitudes
also enter the corresponding scattering cross sections used in the
intranuclear cascade. The nuclear ground state is carried by hole spectral functions
$S(p,E)$---the probability of removing a nucleon of momentum $p$ and leaving
the residual system with excitation energy $E$, normalised separately for
protons and neutrons---obtained for $^{12}$C from correlated-basis function
theory~\cite{Benhar:1994hw, Rocco:2019gfb} and for $^{40}$Ar from $(e,e'p)$
measurements~\cite{JeffersonLabHallA:2022cit, JeffersonLabHallA:2022ljj,
Nikolakopoulos:2024mjj}. Hadrons produced at the vertex are transported
through a sampled nucleon configuration, taken from Green's function Monte
Carlo for carbon~\cite{Carlson:2014vla} and, since argon is not amenable to
continuum quantum Monte Carlo, from single-nucleon densities
there~\cite{Isaacson:2020wlx}; pion absorption follows the Oset
model~\cite{Oset:1987re}. Charged-current, neutral-current and electromagnetic
probes are all supported, so the same code is confronted with $(e,e')$ data as
well as with neutrino measurements, and a tagged hadron beam allows the
cascade to be exercised on its own. \adonis mirrors these choices. The full derivations are
given in the ACHILLES papers~\cite{Isaacson:2020wlx, Isaacson:2022cwh,
Isaacson:2025cnk}; what follows recalls each ingredient only as far as is
needed to show how it is made differentiable, and then describes how gradient
information is carried from the physics parameters through to the final
observables.

Meson-exchange currents and deep-inelastic scattering are absent from ACHILLES and are planned additions to it~\cite{Isaacson:2025cnk}; the same is true of \adonis, and both lie outside the scope of this paper. More generally, the physics choices described above are inherited from ACHILLES rather than selected as preferred models: their purpose here is to provide a realistic generator chain on which to demonstrate differentiability. 
In particular, reproducing the fully stochastic ACHILLES intranuclear cascade
in \adonis demonstrates that differentiability can be retained even through
discrete transport histories.

JAX~\cite{jax2018github} supplies the automatic differentiation, just-in-time
compilation and vectorisation, and allows the same implementation to run on
CPU or GPU. Their relative performance depends strongly on the stage of event
generation. For neutrino interactions on carbon with energies sampled from the
T2K flux, pre-FSI generation showed no GPU acceleration, running at about
$1.8\times10^{3}$ events/s on both a single AMD EPYC CPU core and an NVIDIA
A100 GPU.
The computational bottleneck is instead the intranuclear cascade, whose
throughput increases from about $55$ events/s on a single CPU core to
$560$ events/s on an RTX 2080 Ti GPU and $720$ events/s on an A100 GPU.  A GPU is therefore not required for event generation, but can substantially
accelerate the cascade calculation. The computational performance of the
differentiable inference is discussed separately in
Sec.~\ref{sec:fit_performance}.

We denote by $\bm\theta$ the vector of differentiable physics parameters. For
illustration, in this paper we consider $28$ parameters, chosen to exercise
every part of the generator and to demonstrate operation in high-dimensional
spaces.


\subsection{Differentiability by reweighting}
\label{sec:adonis_reweighting}

A neutrino event generator is an intrinsically stochastic program: it samples a
neutrino energy from a flux, an interaction channel, a struck nucleon from the
nuclear ground state, a hard-scattering kinematic configuration from a squared
amplitude, where applicable the hadronisation of the produced partonic state,
and final-state interactions as the outgoing particles propagate through the
nuclear medium. Differentiating such a program directly is obstructed at every
one of these steps, both by the discrete choices and by the expensive,
tabulated amplitude evaluations that are not themselves written to be
differentiated.

\adonis circumvents this by separating the physics that has to be
differentiated from the physics that does not: the expensive,
well-understood calculation is evaluated \emph{once}, when
the events are generated, and captured in a compact per-event record;
differentiability is then arranged only over the physics parameters of interest,
through a closed-form \emph{reweighting} of that record. Concretely, \adonis
generates a fixed sample of events, a \emph{bank}, and stores with each event the
auxiliary quantities from which its weight can be recomputed as a function of
$\bm\theta$. The generation is a conventional, stochastic forward pass; the object
that is made differentiable is the per-event weight
\begin{equation}
\label{eq:adonis_weight}
w_i(\bm\theta) \;=\; w_i^{0}\,
\underbrace{r^{\rm vertex}_i(\bm\theta)}_{\text{amplitude}}\,
\underbrace{r^{\rm FSI}_i(\bm\theta)}_{\text{cascade}}\,
\underbrace{r^{\rm SF}_i(\bm\theta)}_{\text{nucleus}},
\end{equation}
where $w_i^{0}$ is the fixed proposal weight assigned at generation and each
factor $r_i$ is a differentiable ratio, equal to one at the nominal tune, that
transports the event from the parameters it was generated with to the parameters
$\bm\theta$. The three factors correspond to the three physics blocks above and
are described in the subsections that follow. Because every $r_i$ is a pure JAX
function of $\bm\theta$, so is the product, and the entire map
$\bm\theta \mapsto w_i(\bm\theta)$ is available to automatic differentiation.
Generation and reweighting are performed in double precision (64-bit), since
small variations of the reweighting factors around unity can fall below
single-precision resolution.

Each of the three factors is controlled by its own group of parameters, and
each reweight below is constructed to be identically unity when its parameters
sit at nominal, so that the nominal prediction is untouched. The $28$
parameters divide into eleven acting at the electroweak vertex, eleven in the
cascade, four on the spectral function, and two overall channel
normalisations; each is introduced below in the subsection where it acts.

Each parameter also carries a physically allowed range, which is enforced
wherever it is varied. Most multiplicative rate and normalisation parameters
have nominal value unity and are restricted to positive values; for the
cascade rates of Sec.~\ref{sec:adonis_cascade}, positivity is additionally
required because the total-rate scale $g$ appears in the denominator of
$\exp(-a/g)$. $f^{\rm cex}_{NN}$ is a probability and is restricted to
$[0,1]$. The axial masses enter only through $M_A^2$, so the prediction is
exactly invariant under $M_A \to -M_A$ and the two signs describe the same
physical model. The one boundary that directly affects the gradient, the wall
in $\Delta E_b$, is discussed in Sec.~\ref{sec:adonis_nucleus}.

\subsection{Nuclear ground-state reweighting}
\label{sec:adonis_nucleus}

The target nucleus is described by a spectral function, the probability density
for finding the struck nucleon with momentum $|\mathbf{p}|$ and removal energy
$E$~\cite{Isaacson:2020wlx}. At generation, the struck nucleon is sampled from
the tabulated spectral function by importance sampling, with density proportional to
$|\mathbf{p}|^2 S(|\mathbf{p}|,E)$, and its momentum and removal energy
$E = m_N - E_{\rm in}$ are recorded. Because that point is stored, a deformation
of $S$ reweights the event by the ratio of the deformed to the original density
there, exactly and without resampling,
\begin{align}
r^{\rm SF}_i(\bm\theta) \;=\;\; & N_{\rm SF}
\left[\,1 + (N_{\rm SRC}-1)\,
\sigma\!\left(\frac{|\mathbf{p}|-p_{\rm src}}{w_{\rm src}}\right)\right]
\nonumber\\[2pt]
&\times\;
\frac{S\!\left(|\mathbf{p}|/k_F,\; E - \Delta E_b\right)}
     {S\!\left(|\mathbf{p}|,\; E\right)},
\label{eq:adonis_sf}
\end{align}
where $\sigma$ is the logistic function, $p_{\rm src} = 300$~MeV,
$w_{\rm src} = 80$~MeV, and $(|\mathbf{p}|,E)$ are the values stored for event
$i$. Because the factor is evaluated pointwise on that stored point, it is
differentiable with respect to all four ground-state parameters without
re-sampling the nucleus.

The four parameters act on different features of the same distribution. $k_F$
scales the momentum axis, widening or narrowing the Fermi motion of the struck
nucleon. $\Delta E_b$ shifts the removal-energy axis, and so moves the
position of the quasi-elastic peak; it is the parameter most directly tied to
the reconstructed neutrino energy. $N_{\rm SF}$ is an overall normalisation.
$N_{\rm SRC}$ raises or lowers the region above $p_{\rm src}$, where the
high-momentum tail generated by short-range correlations lives; those
correlations are carried by the tabulated spectral function itself, and
$N_{\rm SRC}$ deforms the tail rather than modelling it.

Two properties of the construction deserve particular attention. The first is that
$\Delta E_b$ is one-sided, and illustrates a subtlety common to any physically
bounded parameter. Negative values are clamped to zero, because the ratio has
a corner where the shifted and unshifted grids part company; once the shift
would push $E$ below zero the model is flat, with an identically vanishing
gradient. We anchor the nominal value on the smooth, one-sided branch
of the response, an offset of $10^{-2}$~MeV from the wall, so that a well-defined
one-sided gradient is always available to the optimiser; the resulting
truncated, non-Gaussian posterior at the boundary is studied explicitly
later in Sec.~\ref{sec:fitting}. The second is that the interpolation of $S$
is cubic in momentum and linear in removal energy, mirroring ACHILLES;
smoother interpolation in removal energy overshoots the steep low-$E$ rise,
producing negative densities.

\subsection{Hard-scattering amplitudes}
\label{sec:adonis_amplitudes}

At the interaction vertex the event is weighted by the squared electroweak
amplitude for the selected channel: quasi-elastic scattering, governed by the
nucleon axial and vector form factors, and resonant pion production, described by
the dynamical coupled-channels amplitudes used in
ACHILLES~\cite{Isaacson:2022cwh, Nakamura:2015rta}. Evaluating these amplitudes
is computationally expensive and, in the case of the tabulated
coupled-channels amplitudes, is not written to be differentiated. Rather than
autodifferentiate through that evaluation, \adonis captures its dependence on the
parameters of interest at generation time.

\emph{The reweighting construction.} The hadron current is linear in the
quantities through which the physics parameters enter,
\begin{equation}
H_b \;=\; \sum_i F_i\,H_{i,b},
\end{equation}
where $b$ labels the nucleon spin combination, the $H_{i,b}$ are unit currents
fixed by the event kinematics and spin state, and the $F_i$ encode the parameter
dependence: form factors in the quasi-elastic case, and more generally form
factors and other amplitude strengths in the resonant case. The squared
amplitude is therefore an exact quadratic form in the $F_i$,
\begin{equation}
|\mathcal{M}|^2
=
\sum_{ij} F_i F_j\, M_{ij},
\quad
M_{ij}
=
\mathrm{Re}\!\sum_{ab}
(L_a\!\cdot\!H_{i,b})^{*}
(L_a\!\cdot\!H_{j,b}),
\label{eq:adonis_quad}
\end{equation}
where $L_a$ is the leptonic current for lepton spin combination $a$, and $a$
and $b$ run over the lepton and nucleon spin combinations, respectively. The
symmetric matrix $M_{ij}$ depends only on the kinematics, not on the physics
parameters, so it is assembled once at generation and stored per event. The
vertex factor of Eq.~\ref{eq:adonis_weight} is then
\begin{equation}
r^{\rm vertex}_i(\bm\theta)
=
\frac{F(\bm\theta)^{\!\top} M\,F(\bm\theta)}
     {F(\bm\theta_0)^{\!\top} M\,F(\bm\theta_0)},
\label{eq:adonis_vertex}
\end{equation}
where $\bm\theta_0$ denotes the nominal values. This is an identity rather than
a fit: the underlying amplitude calculation is never rerun after generation,
the weight is exactly unity at nominal, and because every parameter enters only
through the smooth map $\bm\theta \mapsto F$, the construction is exact for
arbitrary simultaneous variations and differentiable to any order. For
quasi-elastic scattering, $F=\{F_1,F_2,F_A,F_P\}$, so $M$ is $4\times4$; the
resonant channel is treated analogously, with separate vector, axial and
pion-pole contributions for each partial wave.

Where the parameters of interest affect only a subset of these contributions,
the quadratic separates exactly. Splitting $F$ into the components that vary,
$F_{\rm var}$, and those held at their nominal values, $F_{\rm fix}$,
\begin{equation}
|\mathcal{M}|^2
=
F_{\rm var}^{\!\top} M_{\rm vv} F_{\rm var}
+ 2\,F_{\rm var}^{\!\top} v + c,
\end{equation}
with $v = M_{\rm vf} F_{\rm fix}^0$ and
$c = F_{\rm fix}^{0\top} M_{\rm ff} F_{\rm fix}^0$, so that only
$(M_{\rm vv},v,c)$ need be stored. The fixed contributions are compressed into
$v$ and $c$ rather than discarded, and the record grows or shrinks with the
list of parameters exposed.

\emph{The quasi-elastic current.} Quasi-elastic scattering proceeds through
the one-body weak current. For charged-current scattering it takes the standard
form~\cite{LlewellynSmith:1971uhs},
\begin{align}
J^\mu = \bar{u}(p')\Big[\,& F_1\gamma^\mu
+ \frac{i F_2}{2m_N}\sigma^{\mu\nu}q_\nu \nonumber\\
&+ F_A\gamma^\mu\gamma_5
+ \frac{F_P}{m_N}q^\mu\gamma_5 \Big] u(p),
\label{eq:adonis_ls}
\end{align}
with $q$ the four-momentum transfer. Neutral-current elastic scattering has
the same four current structures, so the quadratic construction of
Eq.~\ref{eq:adonis_quad} applies unchanged; only the electroweak combinations
entering their coefficients differ. In particular, the vector contributions
depend on different proton and neutron form-factor combinations, while the
axial contribution changes sign between proton and neutron. Strange form
factors are not included in ACHILLES and are therefore also absent from
\adonis.

The vector form factors $F_1$ and $F_2$ describe the charge and magnetisation
distributions of the nucleon and are fixed by electron scattering; $F_A$ and
$F_P$ are the axial and induced pseudoscalar form factors. In addition to the
parameters entering these form factors themselves, \adonis introduces two
overall scaling parameters: $S_V^{\rm QE}$ scales $F_1$ and $F_2$ together,
while $S_A^{\rm QE}$ scales $F_A$ and $F_P$ together.

\emph{Vector form factors.} $F_1$ and $F_2$ are built from the Sachs electric
and magnetic form factors,
\begin{equation}
F_1 = \frac{G_E + \tau G_M}{1+\tau}, \qquad
F_2 = \frac{G_M - G_E}{1+\tau},
\end{equation}
with $\tau = Q^2/4m_N^2$. These are taken from the Kelly
parametrisation~\cite{Kelly:2004hm}, in which each Sachs form factor is a
ratio of polynomials in $\tau$,
\begin{equation}
G(\tau) \;=\; \frac{1 + a_1\tau}{1 + b_1\tau + b_2\tau^2 + b_3\tau^3},
\end{equation}
fitted to elastic electron--proton data and constructed to fall with the
correct power of $Q^2$ at large momentum transfer. The four parameters
$\mu_p$, $\mu_n$, $G_E^p$ and $G_E^n$ scale $G_M^p$, $G_M^n$, $G_E^p$ and
$G_E^n$ individually, and enter Eq.~\ref{eq:adonis_vertex} through the linear
relations above.

\emph{Axial form factor.} The axial form factor is taken as a $z$-expansion,
the model-independent parametrisation that maps the analytic domain of
$F_A(Q^2)$ onto the unit disc,
\begin{equation}
z(Q^2) = \frac{\sqrt{t_c+Q^2}-\sqrt{t_c-t_0}}
              {\sqrt{t_c+Q^2}+\sqrt{t_c-t_0}}, \qquad
F_A = \sum_{k} a_k\,z^k,
\end{equation}
with $t_c = 9m_\pi^2$ the three-pion production threshold and $t_0$ a fixed
expansion point. The coefficients are those of Ref.~\cite{Meyer:2016oeg},
obtained from neutrino--deuteron data. Because this form has no axial mass,
$M_A^{\rm QE}$ acts instead through the ratio of two dipoles,
\begin{align}
F_A^{\rm dip}(Q^{2};M_A) &= \frac{-g_A}{\left(1+Q^{2}/M_A^{2}\right)^{2}},
\nonumber\\
r(Q^{2};M_A) &= \frac{F_A^{\rm dip}(Q^{2};M_A)}
                     {F_A^{\rm dip}(Q^{2};1.0~{\rm GeV})},
\end{align}
evaluated at the momentum transfer $Q^2$ of each event, which is stored in the
event record, and multiplying the axial block in
Eq.~\ref{eq:adonis_vertex}. It therefore deforms the axial $Q^2$ dependence
about the nominal in the direction an axial mass would, and is unity at
$M_A=1.0$~GeV by construction. Only $M_A^2$ appears, so the ratio is even in
$M_A$. The induced pseudoscalar follows from $F_A$ by the pion-pole dominance
relation $F_P = 2m_N^2 F_A/(Q^2+m_\pi^2)$ and is therefore carried along by
both $S_A^{\rm QE}$ and $M_A^{\rm QE}$.

\emph{Resonant vertex.} Single-pion production is described by the ANL--Osaka
dynamical coupled-channels model~\cite{Nakamura:2015rta}. Rather than summing
independent Breit--Wigner resonances, it solves coupled scattering equations
for the meson--baryon channels $\pi N$, $\eta N$, $K\Lambda$ and $K\Sigma$
simultaneously, so that resonant and non-resonant contributions and their
interference emerge from one unitary amplitude. Only $\pi N$ is produced at
the interaction vertex; the remaining channels enter through the coupled-channel
solution rather than as generated final states. The
amplitude is decomposed into partial waves labelled $L_{2I\,2J}$ by the
orbital angular momentum, isospin and total angular momentum of the
pion--nucleon pair; fourteen are tabulated,
\begin{equation*}
\begin{aligned}
&S_{11},\; S_{31},\; P_{11},\; P_{13},\; P_{31},\; P_{33},\; D_{13},\\
&D_{15},\; D_{33},\; D_{35},\; F_{15},\; F_{17},\; F_{35},\; F_{37},
\end{aligned}
\end{equation*}
each containing the resonances with the corresponding quantum numbers. All
fourteen contribute to the nominal \adonis prediction, and each can carry an
independent differentiable normalisation because the amplitude is linear in
that contribution. For simplicity, the studies below vary only $S_\Delta$, which scales the
$P_{33}$ partial wave that contains the $\Delta(1232)$ resonance and dominates
pion production at accelerator energies.

Beyond the partial-wave normalisations, $F_{\rm pp}$ scales the longitudinal
pion-pole term of the charged-current resonant current, analogous to the
induced pseudoscalar contribution discussed above. The axial current is
controlled by $C_5^A$ and $M_A^{\rm RES}$: the former scales its overall
strength, while the latter modifies its $Q^2$ dependence using the same
dipole-ratio construction introduced for $M_A^{\rm QE}$.

Neutral-current resonant production uses the same DCC decomposition. The
partial-wave normalisations and the axial parameters $C_5^A$ and
$M_A^{\rm RES}$ are common to charged- and neutral-current production. The
differences lie in the electroweak current: the vector contribution uses the
corresponding neutral-current combination of isovector and isoscalar
components, and the pion-pole term is absent. Consequently, $F_{\rm pp}$
affects only charged-current resonant events.

In this way, the event-dependent ingredients of the amplitude are evaluated
once at generation, while its dependence on the axial masses, form-factor
scales and strengths remains explicit and can be differentiated exactly at
negligible fit-time cost.

\subsection{Intranuclear cascade}
\label{sec:adonis_cascade}

The hadrons produced at the vertex propagate through the residual nucleus and
undergo final-state interactions, modelled with a JAX re-implementation of the intranuclear cascade used in
ACHILLES~\cite{Isaacson:2020wlx}: elastic and inelastic nucleon--nucleon and
pion--nucleon scattering, pion absorption, and charge exchange. Everything at
the vertex is a smooth function of the parameters once the kinematics are
fixed. The cascade is not: a hadron leaving the vertex takes a variable-length
sequence of sampled re-interactions, and the number, position and identity of
those interactions are themselves random. Differentiating any single realised
history gives nothing, because a parameter change does not move that history,
it changes how probable it was.

\emph{The construction.} \adonis compactly stores the history each
event actually took, and recomputes its probability under new parameters. The
cascade factor in Eq.~\ref{eq:adonis_weight} is the likelihood ratio of that
history,
\begin{equation}
r^{\rm FSI}_i(\bm\theta) = \prod_{n\,\in\,{\rm history}_i}
\frac{p_n(\bm\theta)}{p_n({\rm nominal})},
\label{eq:adonis_fsi}
\end{equation}
a product over the interaction and survival probabilities of every step, each a
smooth function of the corresponding rate parameter and equal to unity at
nominal. Reweighting the stored history in this way is the importance-weight
expression of the change of rates, and it is differentiable in the cascade
parameters without re-running the cascade. It is closely related to the
score-function construction for stochastic computation
graphs~\cite{Foerster:2018dice}: there the derivative of an expectation is
obtained from the log-probability of the realised choices, whereas here the
full likelihood ratio of each stored history is written down directly. The
finite parameter dependence is therefore available explicitly, and its
derivatives at any $\bm\theta$ follow by automatic differentiation.

\emph{The step probabilities.} A propagating hadron advances in small steps.
At each step the nucleons within reach are candidates, and the hadron
interacts with the closest of them with probability
\begin{equation}
p_0 = \exp(-a), \qquad a = \frac{\pi b^{2}}{\sigma_{\rm tot}},
\end{equation}
where $b$ is the impact parameter to that nucleon and $\sigma_{\rm tot}$ the
total cross section there. A rate parameter enters through
$\sigma_{\rm tot}$ and so has two distinct effects. First, it changes
\emph{whether} the hadron interacts: rescaling the total by a factor $g$ maps
$a \to a/g$ and gives an interaction-probability factor $p_g/p_0$ on a step
that interacted and a survival-probability factor $(1-p_g)/(1-p_0)$ on one
that did not, with $p_g = \exp(-a/g)$. Second, it changes \emph{which} channel
occurred, contributing $s_{\rm realised}/g$ on a step that interacted. A
common rescaling of every channel leaves the second factor at unity but not
the first, which is the mean-free-path effect.

\emph{Pion channels.} A pion may scatter elastically, charge exchange, be
absorbed on a nucleon pair, or convert to $\eta N$; the four parameters
$s^{\rm el}_{\pi N}$, $s^{\rm cex}_{\pi N}$, $s_{\rm abs}^{\pi}$ and
$s_{\rm conv}$ scale those four rates respectively, and
\begin{align}
g \;=\; 1
&+ (s_{\rm abs}^{\pi}-1)\,f_{\rm abs}
+ (s^{\rm el}_{\pi N}-1)\,f_{\rm el}
\nonumber\\
&+ (s^{\rm cex}_{\pi N}-1)\,f_{\rm cex}
+ (s_{\rm conv}-1)\,f_{\rm conv},
\end{align}
with $f_i$ the fraction of the total cross section carried by channel $i$ at
that step. Written in this form $g$ is exactly unity at nominal in any
precision, which matters because $\exp(-a/g)$ amplifies any error in $g$. A
pion converted through $\pi N\to\eta N$ produces a propagating $\eta$, which
may subsequently back-convert to a pion or leave the nucleus; the
strangeness-producing branch is treated as terminal. The underlying cross
sections are those of ACHILLES: the ANL--Osaka amplitudes of
Sec.~\ref{sec:adonis_amplitudes} for meson--baryon
scattering~\cite{Nakamura:2015rta}, and the optical-potential model of Oset and
Salcedo for absorption~\cite{Oset:1987re}, in which the pion is absorbed
through the imaginary part of its self-energy in nuclear matter.

\emph{Nucleon channels.} A nucleon may scatter elastically or inelastically,
the latter proceeding through $NN \to N\Delta$ with the $\Delta$ decaying
immediately to $N\pi$; this is the channel that degrades fast nucleons and
creates pions inside the nucleus. Both rates are resolved by isospin pair,
since $pp$, $pn$ and $nn$ cross sections differ, giving six parameters
$s^{\rm el}_{NN,ij}$ and $s^{\rm inel}_{NN,ij}$ that act only on steps of the
matching pair, with
$g = s^{\rm el}(1-f_{\rm inel}) + s^{\rm inel} f_{\rm inel}$. The elastic
cross section follows the GiBUU parametrisation~\cite{Buss:2011mx} and the
inelastic one the $NN \to N\Delta$ model of Dmitriev and
Sushkov~\cite{Dmitriev:1986st}.

\emph{Charge exchange.} The last cascade parameter is not a rate but a
probability. When a proton and a neutron scatter elastically the outgoing pair
may or may not exchange identity, and ACHILLES assigns the two outcomes equal
weight. $f^{\rm cex}_{NN}$ varies that split, contributing
$f^{\rm cex}_{NN}/0.5$ on a step that exchanged and
$(1-f^{\rm cex}_{NN})/0.5$ on one that did not. Identical pairs cannot
exchange, so $pp$ and $nn$ steps are unaffected and carry no factor. Its
nominal value is $0.5$ rather than unity because it is a fraction, not a
scale.

\subsection{Channel normalisations}
\label{sec:adonis_norms}

The last two parameters, $N_{\rm QE}$ and $N_{\rm RES}$, multiply the weights
of the quasi-elastic and resonant samples as a whole. They therefore change no
within-channel kinematic shape, but control the relative rates of the two
interaction channels.

\subsection{Differentiable events and observables}
\label{sec:adonis_events}

The result of the generation is a bank in which every event carries, alongside
its final-state kinematics, the records of
Secs.~\ref{sec:adonis_nucleus}--\ref{sec:adonis_cascade} that render its weight
$w_i(\bm\theta)$ of Eq.~\ref{eq:adonis_weight} an exact, differentiable function
of the physics parameters. Differentiability is thus a property of each stored
event individually: for any event one can ask, and answer exactly, how its weight
responds to any physics parameter.

This event-by-event differentiability is what makes the downstream analysis
differentiable for free. Physics observables are built from the events by
operations that do not depend on $\bm\theta$: a signal or topology definition
selects events by their final state, a choice of kinematic variable maps each
event to an axis value, and a binning assigns each event to a bin. None of these
touches the weights, so a binned prediction is simply the sum of the differentiable
per-event weights falling in each bin,
\begin{equation}
\label{eq:adonis_binned}
m_b(\bm\theta) = \sum_{i \in b} w_i(\bm\theta),
\qquad
\frac{\partial m_b}{\partial\bm\theta} = \sum_{i \in b}
\frac{\partial w_i}{\partial\bm\theta}.
\end{equation}
The gradient of any histogram, and therefore of any differentiable downstream
statistic built from it, is obtained by propagating the gradients of the
weights through the same selection and binning. Changing the signal definition, the observable, or the bin
edges changes which events enter each sum but not the differentiability of the
result. In sum, an analysis is free to define its observables however it wishes
and retains exact gradients with respect to the physics parameters throughout.

Nothing in Eq.~\ref{eq:adonis_binned} requires an event to belong to a single
bin. The sum may equally be taken over a weight $\alpha_{ib}$ that spreads each
event across bins, $m_b = \sum_i \alpha_{ib}\,w_i(\bm\theta)$, and the
gradient follows unchanged. When an event's observable is known only to within
a reconstruction uncertainty, that uncertainty can therefore be carried into
the prediction as a soft assignment rather than collapsed into a hard choice
of bin.

\subsection{Gradients and higher-order derivatives}
\label{sec:adonis_gradients}

Because $\bm\theta \mapsto \{w_i\}$ is a JAX function, the Jacobian of the binned
prediction,
\begin{equation}
J_{bk} = \frac{\partial m_b}{\partial\theta_k},
\end{equation}
is obtained by automatic differentiation. As the map takes few inputs (the
parameter vector) to many outputs (the per-event weights, reduced to bins), we
evaluate it in forward mode. A single Jacobian-vector product with a unit tangent
$\mathbf{e}_k$ returns the full per-event derivative in the $k$-th parameter
direction in one forward pass, and the columns for all parameters are batched into
one vectorised evaluation whose cost is a small multiple of a single forward
reweight.

The same construction extends to arbitrary derivative order by nesting
Jacobian-vector products along the desired parameter directions, with no
symbolic derivative code required. The first derivatives form the
Gauss--Newton, or Fisher, matrix
\begin{equation}
\label{eq:adonis_fisher}
F_{kl} = \sum_b \frac{1}{\sigma_b^{2}}\,
\frac{\partial m_b}{\partial\theta_k}\,
\frac{\partial m_b}{\partial\theta_l},
\end{equation}
with $\sigma_b$ the per-bin uncertainty. This is the positive-semidefinite
curvature approximation underlying the fits of Sec.~\ref{sec:fitting}.

\section{Validation}
\label{sec:validation}

\adonis implements the same underlying physics choices as ACHILLES, enabling a
direct validation against its predictions. Ref.~\cite{Isaacson:2025cnk}
compares ACHILLES with a broad set of measurements; here we reproduce
representative comparisons with \adonis and ACHILLES overlaid, spanning
different probes, interaction channels, target nuclei and beam energies.

\begin{figure}[htbp]\centering
  \includegraphics[width=\linewidth]{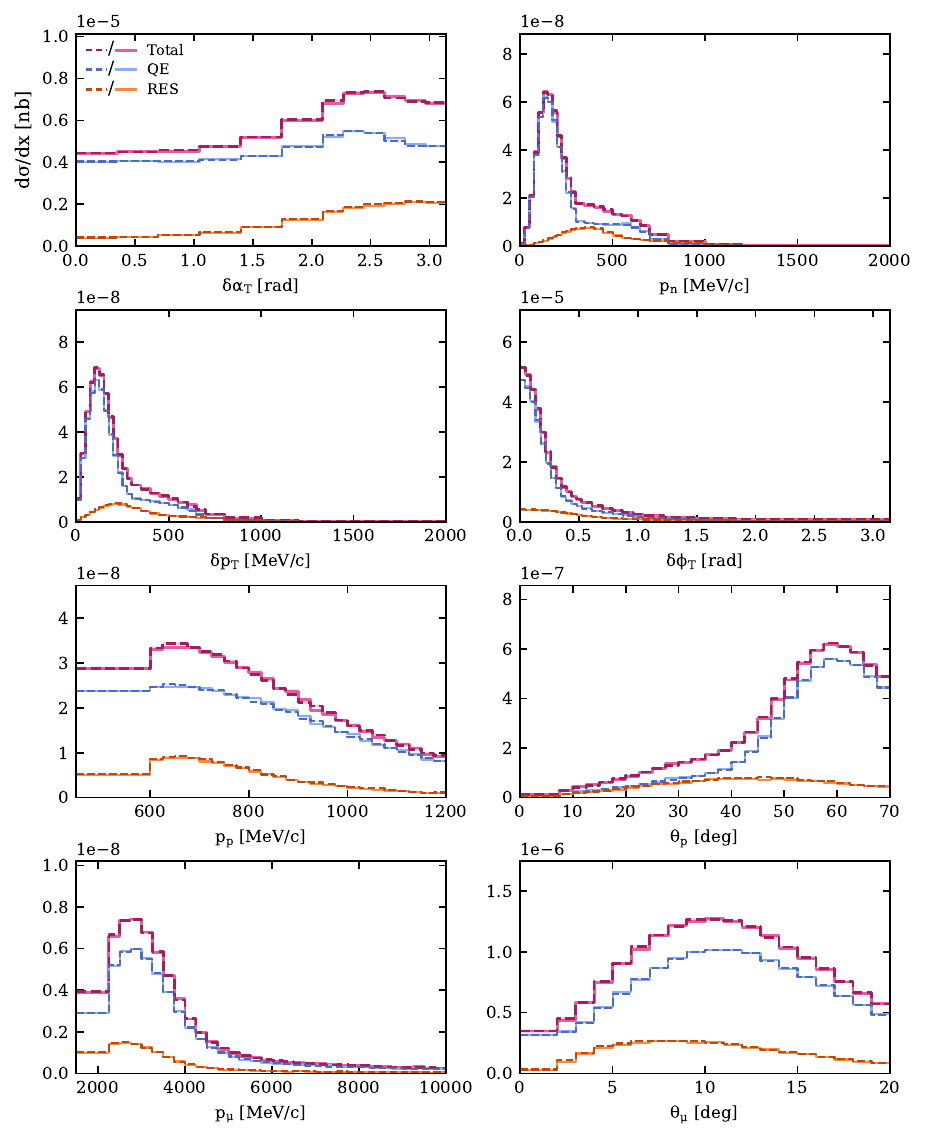}
  \caption{ADoNIS (solid) vs.\ ACHILLES (dashed). MINERvA CC0$\pi$ on $^{12}$C at $\langle E_\nu\rangle \approx 3$~GeV, with the
    signal definition of Ref.~\cite{MINERvA:2018hba}: the TKI
    variables $\delta\alpha_T$, $p_n$, $\delta p_T$ and $\delta\phi_T$, and the proton and
    muon kinematics.}
  \label{fig:val_minerva}
\end{figure}

Figure~\ref{fig:val_minerva} shows the MINERvA CC0$\pi$ sample on $^{12}$C at
$\langle E_\nu\rangle \approx 3$~GeV, generated for the signal definition of
Ref.~\cite{MINERvA:2018hba}. The observables span lepton kinematics, hadron
kinematics and the transverse kinematic imbalance (TKI) variables that
correlate the two. At MINERvA energies both quasi-elastic and resonant interactions
contribute substantially, with excellent agreement observed in both. Agreement
holds equally for other beams, binnings and signal definitions. For instance,
we showcase the T2K CC0$\pi$ and CC1$\pi^+$ samples of
Refs.~\cite{T2K:2018rnz} and~\cite{T2K:2021naz} in
Figs.~\ref{fig:val_t2k_cc0pi} and~\ref{fig:val_t2k_cc1pi} of
Appendix~\ref{app:validation}.

Figure~\ref{fig:val_uboone_nc2d} showcases the double-differential cross
section for NC~1$\pi^0$Xp production measured by
MicroBooNE~\cite{MicroBooNE:2024pdj}, and demonstrates that equivalent
predictions are also achieved for the two generators for neutral-current
observables on Ar. Analogous agreement in charged-current MicroBooNE TKI
observables~\cite{MicroBooNE:2023tzj} is also presented in
Fig.~\ref{fig:val_uboone_cc1p} in Appendix~\ref{app:validation}.

\begin{figure}[htbp]\centering
  \includegraphics[width=\linewidth]{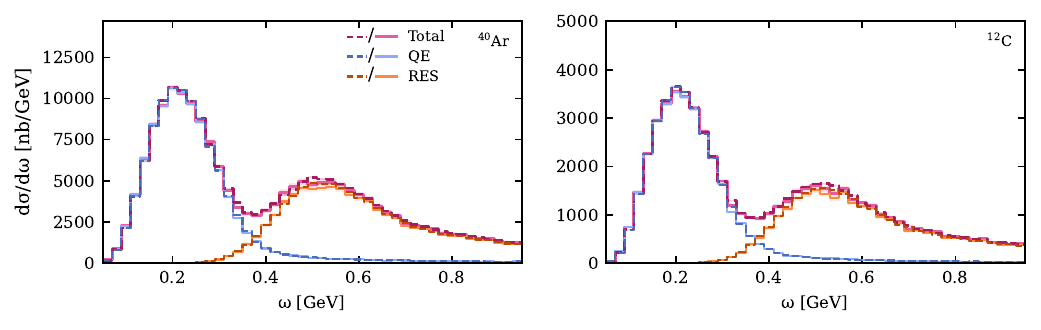}
  \caption{ADoNIS (solid) vs.\ ACHILLES (dashed). Inclusive $(e,e')$ cross section $d\sigma/d\omega$ on $^{12}$C and $^{40}$Ar at a beam energy of $2.222$~GeV and $\theta_{e'}\approx 15.5^\circ$~\cite{Murphy:2019wed, JeffersonLabHallA:2022cit, JeffersonLabHallA:2022ljj}, separated into quasi-elastic and resonant contributions.}
  \label{fig:val_ee}
\end{figure}

The last two comparisons isolate single ingredients. Figure~\ref{fig:val_ee}
shows inclusive $(e,e')$ cross sections on $^{12}$C and $^{40}$Ar at
$2.222$~GeV, in the configuration of the Jefferson Lab E12-14-012
measurements~\cite{Murphy:2019wed, JeffersonLabHallA:2022cit,
JeffersonLabHallA:2022ljj}: an electromagnetic probe of known kinematics, on
both target nuclei, separated into quasi-elastic and resonant contributions.
Exclusive electron-scattering observables from e4$\nu$~\cite{CLAS:2021neh},
which confront the neutrino-energy estimators against a beam of known energy,
are given in Fig.~\ref{fig:val_e4nu} of Appendix~\ref{app:validation}. Lastly, Fig.~\ref{fig:val_pinuc} shows
$\pi^+$--nucleus absorption and reaction cross sections in the configuration
of Refs.~\cite{ASHERY, LADS:2000pzh, DUET}, where the pions come from a tagged
hadron beam rather than an interaction vertex, so that the cascade is
exercised on its own.

Taken together, these comparisons show that the differentiable reformulation
preserves the underlying physics predictions.

\begin{figure}[htbp]\centering
  \includegraphics[width=\linewidth]{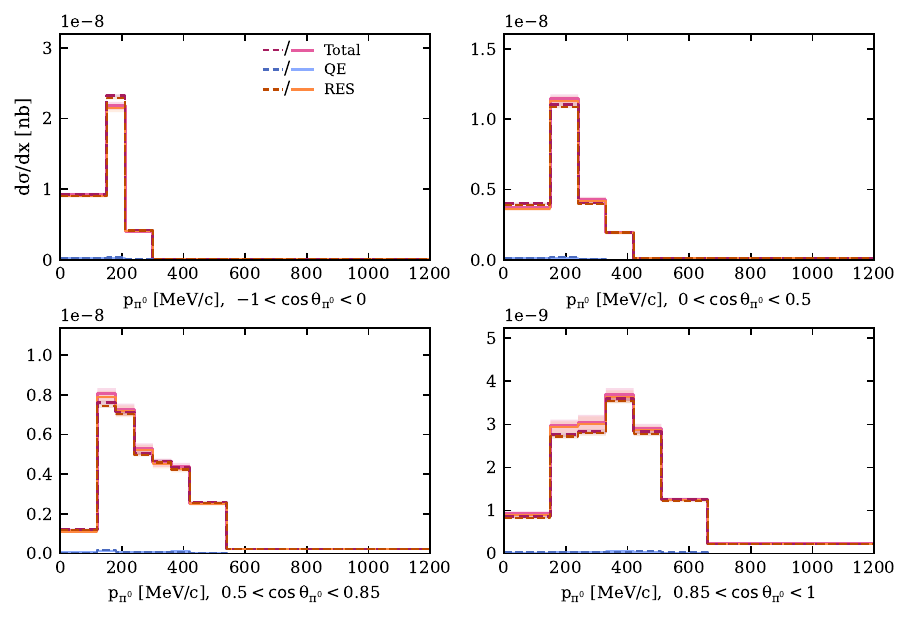}
  \caption{ADoNIS (solid) vs.\ ACHILLES (dashed). MicroBooNE NC~1$\pi^0$Xp on $^{40}$Ar~\cite{MicroBooNE:2024pdj}: the double-differential cross section $d^2\sigma/d\cos\theta_{\pi^0}\,dp_{\pi^0}$, in slices of $\cos\theta_{\pi^0}$.}
  \label{fig:val_uboone_nc2d}
\end{figure}

\begin{figure}[htbp]\centering
  \includegraphics[width=\linewidth]{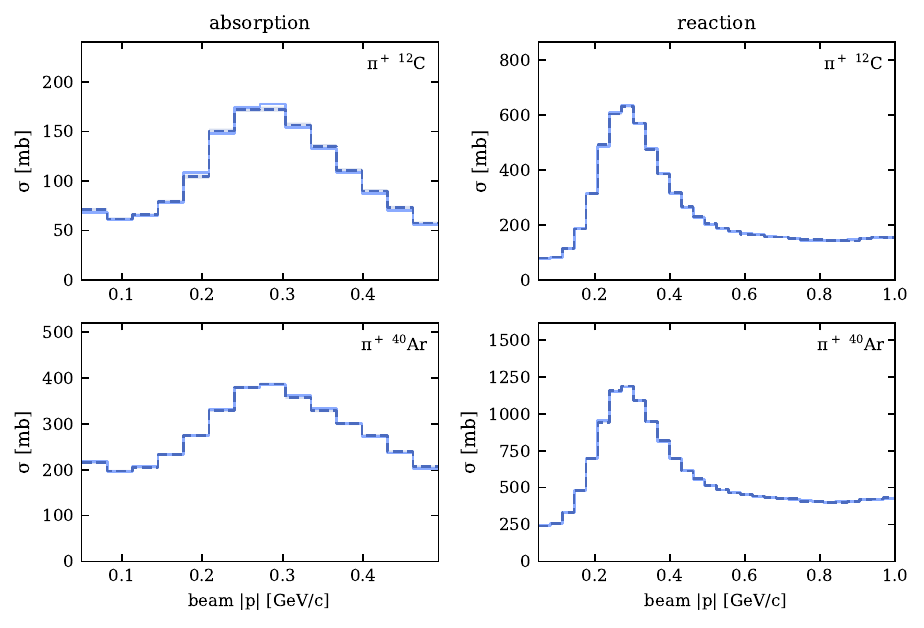}
    \caption{ADoNIS (solid) vs.\ ACHILLES (dashed). $\pi^+$--nucleus absorption
    and reaction cross sections on carbon and argon as a function of pion
    momentum~\cite{ASHERY, LADS:2000pzh, DUET}, testing the intranuclear cascade
    in isolation.}
    \label{fig:val_pinuc}
\end{figure}

\section{Gradient information}
\label{sec:jacobian}

\begin{figure*}[t]\centering
  \includegraphics[width=\linewidth]{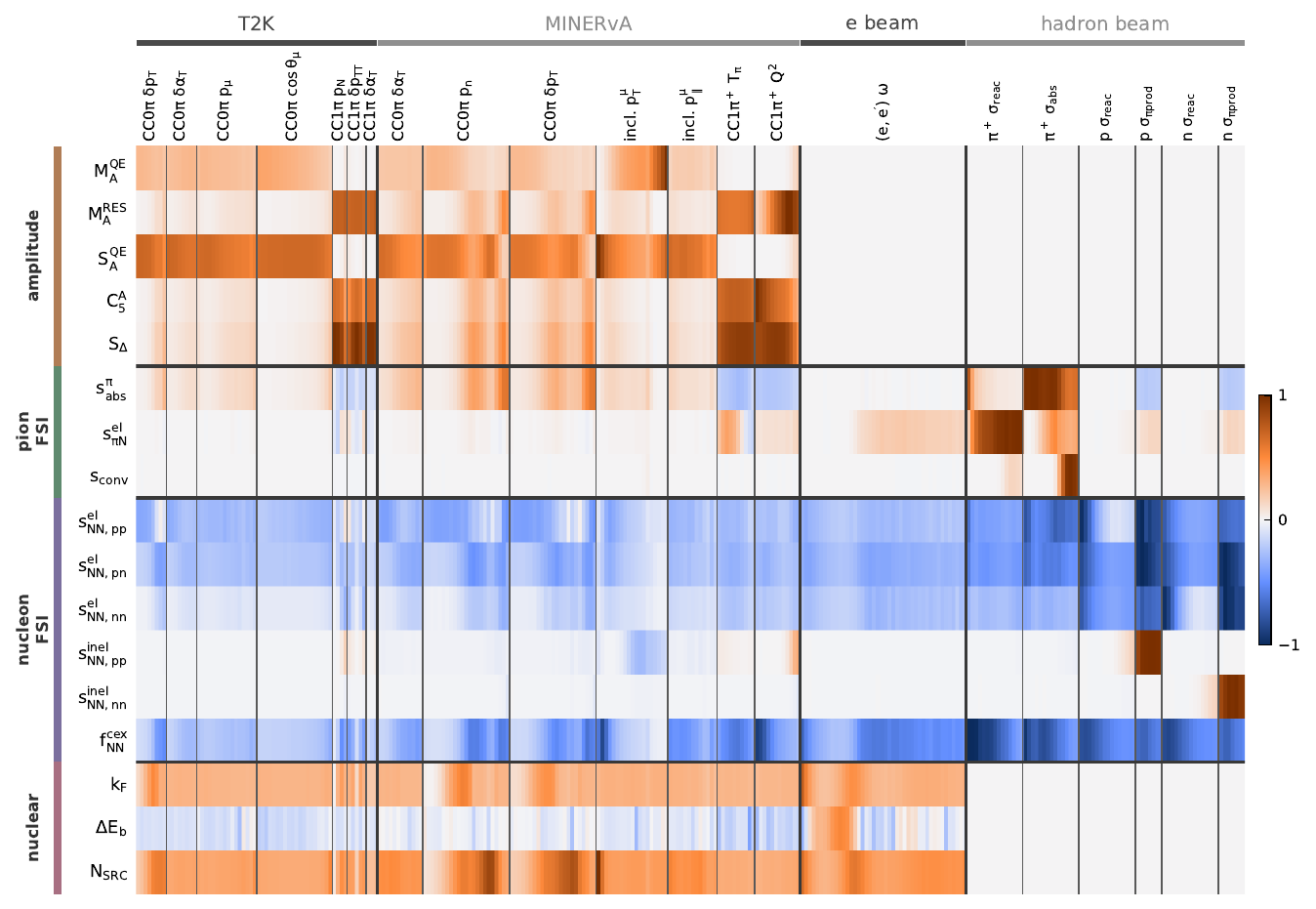}
  \caption{Signed per-bin parameter response
  $\delta\theta_k J_{ik}/\sigma_i$, with each row normalised to its own peak,
  for the $17$ parameters retained for inference and highlighted in
  Fig.~\ref{fig:probe_subsets}, across the samples and fitted bins shown in
  Fig.~\ref{fig:fit_rates}.}
  \label{fig:gradient_shape}
\end{figure*}

Neutrino--nucleus scattering is being probed by an increasingly broad set of
measurements. The Fermilab short-baseline
programme~\cite{MicroBooNE:2015bmn} now has all three of its liquid-argon
detectors operating---MicroBooNE~\cite{MicroBooNE:2016pwy},
ICARUS~\cite{ICARUS:2023gpo} and SBND~\cite{SBND:2025lha}---providing a growing
body of measurements of neutrino--argon
scattering~\cite{MicroBooNE:2024bnl, MicroBooNE:2024yzp, MicroBooNE:2024zwf,
MicroBooNE:2024pdj, MicroBooNE:2025aiw, MicroBooNE:2025pvb, ICARUS:2026nbg}.
NOvA~\cite{NOvA:2004blv} and T2K~\cite{T2K:2011qtm} continue to take data and
report cross sections on carbon, and in the case of T2K on oxygen as
well~\cite{NOvA:2024rov, NOvA:2024zmr, NOvA:2026zup, T2K:2025kdk, T2K:2025smz,
T2K:2026rru}. MINERvA, although no longer running, has released its data
publicly~\cite{MINERvA:2025cfj} and continues to produce
measurements~\cite{MINERvA:2025tem, MINERvA:2025hzq, MINERvA:2026apf,
MINERvA:2026ymp, MINERvA:2026naw}. The upgraded T2K near
detector~\cite{T2K:2019bbb} is complemented on the same beamline by
WAGASCI--BabyMIND~\cite{T2K:2025kda} and NINJA~\cite{NINJA:2020gbg}.
Future facilities will further broaden this programme, including the
intermediate water Cherenkov detector and further near-detector upgrades
proposed for Hyper-Kamiokande~\cite{Hyper-Kamiokande:2018ofw,
Hyper-Kamiokande:2025asb}, and the DUNE near-detector complex, including SAND
and the movable PRISM configuration~\cite{DUNE:2021tad, DUNE:2025lvs}.
Complementary measurements constrain the same nuclear physics with
better-controlled probes, including electron scattering at Jefferson Lab and
MAMI~\cite{Murphy:2019wed, Blomqvist:1998xn}, the continuing e4$\nu$
programme~\cite{CLAS:2021neh, CLAS:2025fqh}, and hadron--nucleus scattering,
whose existing data are used in cascade tuning~\cite{PinzonGuerra:2018rju}
and are being extended by liquid-argon test-beam
programmes~\cite{DUNE:2020cqd}.

Taken together, these data provide overlapping constraints on the same
underlying interaction and nuclear-model parameters, but their complementarity
is not always apparent from the predictions themselves. In conventional
workflows, questions such as which measurement constrains which parameter,
where different datasets provide redundant or complementary information, and
which parameter combinations remain degenerate can only be answered through
dedicated fits and sensitivity studies. Differentiability makes this
information directly accessible through the response of the prediction to its
parameters.

To demonstrate this, we assemble the Jacobian of
Sec.~\ref{sec:adonis_gradients} across every bin of a concrete set of
measurements and ask what each one constrains. The samples are chosen to span
the kinds of information available rather than to be exhaustive. Each mirrors
the signal definition and, where one is published, the binning of a specific
measurement. From T2K we take the CC0$\pi$ distributions in $\delta p_T$,
$\delta\alpha_T$, $p_\mu$ and $\cos\theta_\mu$~\cite{T2K:2018rnz} and the
CC1$\pi^+$ distributions in $p_N$, $\delta p_{TT}$ and
$\delta\alpha_T$~\cite{T2K:2021naz}; from MINERvA the CC0$\pi$ distributions
in $\delta\alpha_T$, $p_n$ and $\delta p_T$~\cite{MINERvA:2018hba}, the
inclusive muon $p_T$ and $p_\parallel$~\cite{MINERvA:2018hqn}, and the
CC1$\pi^+$ distributions in $T_\pi$ and $Q^2$~\cite{MINERvA:2026ymp}.
Electron scattering enters as the inclusive $(e,e')$ energy-transfer spectrum
at $2.222$~GeV and $\theta_{e'} \approx
15.5^\circ$~\cite{Murphy:2019wed, JeffersonLabHallA:2022cit,
JeffersonLabHallA:2022ljj}. Finally, tagged $\pi^+$, $p$
and $n$ beams contribute their reaction and, respectively, absorption and
pion-production cross sections as functions of beam momentum. Between them
these span three probes, two neutrino-beam energies a factor of five apart,
exclusive and inclusive observables, and a purely hadronic probe that involves
no interaction vertex at all. The targets are carbon, except for the three
CC1$\pi^+$ samples, which are on hydrocarbon and so carry a free-hydrogen
contribution, in the gradient as well as in the prediction.

Figure~\ref{fig:gradient_shape} shows where in each observable the $17$
parameters used in Sec.~\ref{sec:fitting} have an effect. We take a reference
variation $\delta\theta_k$ of $20\%$ of the nominal value for each parameter,
except for $\Delta E_b$, whose nominal value is close to zero and for which we
use $4$~MeV. For each bin $i$ we compute the signed response
$\delta\theta_k J_{ik}/\sigma_i$, where $\sigma_i$ is an assumed $5\%$ uncertainty in bin $i$, and normalise each parameter row to its
largest absolute value. The figure therefore shows the shape of each
parameter's response across phase space, rather than its overall strength.

Similar response shapes indicate degeneracies: if two parameters change the
bins in the same way, the data cannot tell which one caused the change. Bins
where their responses differ are therefore the ones that can separate them.
The figure also shows which parts of a distribution constrain each parameter.
Sensitivity across the bulk of a distribution makes inference more robust,
while constraints driven mainly by bins in the distribution tails are more
prone to bias from small mismodellings.

\begin{figure}[htbp]\centering
  \includegraphics[width=\linewidth]{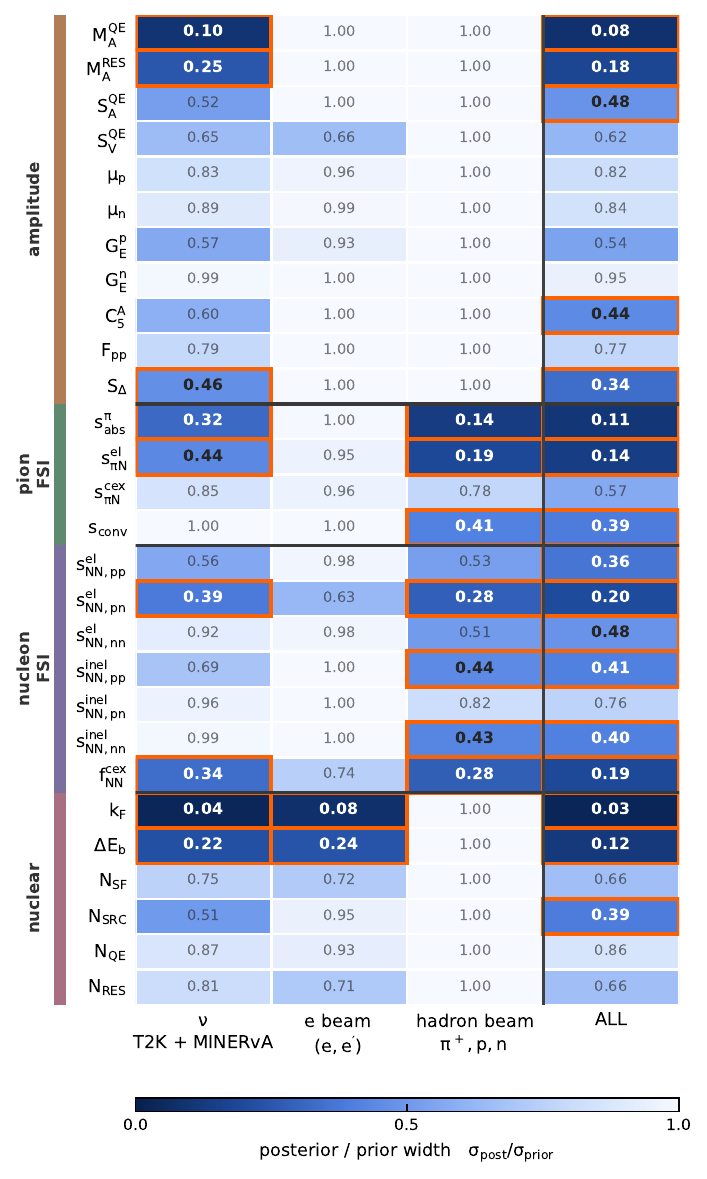}
  \caption{Marginalised parameter constraints
  $\sigma_{\rm post}/\sigma_{\rm prior}$ for samples aggregated by probe type and
  for all samples combined. Cells with
  $\sigma_{\rm post}/\sigma_{\rm prior}<0.5$ are highlighted in orange.}
  \label{fig:probe_subsets}
\end{figure}

Figure~\ref{fig:probe_subsets} uses the same Jacobian to construct the Fisher
information of Eq.~\ref{eq:adonis_fisher}. We assign each parameter a Gaussian
prior with width $\delta\theta_k$, using the same reference variations defined
above, and compute the marginalised posterior-to-prior width
$\sigma_{\rm post}/\sigma_{\rm prior}$ for each probe. A value of one means
that the probe adds essentially no information beyond the prior, while smaller
values indicate a stronger constraint. For the inference studies of
Sec.~\ref{sec:fitting}, we retain the $17$ parameters for which at least one
probe gives $\sigma_{\rm post}/\sigma_{\rm prior}<0.5$. These are highlighted
in orange and are the same parameters shown in
Fig.~\ref{fig:gradient_shape}.

The pattern has a direct physical interpretation. The hadron-beam column leaves
every vertex and ground-state parameter exactly at its prior and constrains only
the cascade, as expected since no interaction vertex is involved. It is
nonetheless the most informative probe of nucleon and pion final-state
interactions. The neutrino and electron samples together constrain the nuclear
parameters, while the neutrino samples alone are the most informative about the
form factors entering the amplitude. Parameter correlations are equally
visible. $C_5^A$, for instance, is directly constrained only by the neutrino
samples. Adding the electron and hadron beams, neither of which is directly
sensitive to it, nevertheless tightens its constraint by constraining the
parameters with which it is correlated.

Three uses follow directly. The first is fit design: the gradients show which
parameters the data can constrain and which samples add new information beyond
those already included. The second is measurement design. Multi-dimensional
differential cross sections are already being
published~\cite{MicroBooNE:2023foc, NOvA:2026zup, MINERvA:2026naw}, and the
statistics of the next generation will support increasingly fine and
high-dimensional measurements. Because observables and binnings can be defined
after the event-level gradients are available, alternative measurements can be
compared directly by the parameter information they provide, allowing the
observables and binnings that best constrain the parameters of interest to be
identified. The third is experiment design, where different beam energies,
targets, probe types and final-state capabilities can be compared by the
additional information they provide.

\section{Parameter Inference and Uncertainty Quantification}
\label{sec:fitting}

\begin{figure*}[t]\centering
  \includegraphics[width=\linewidth]{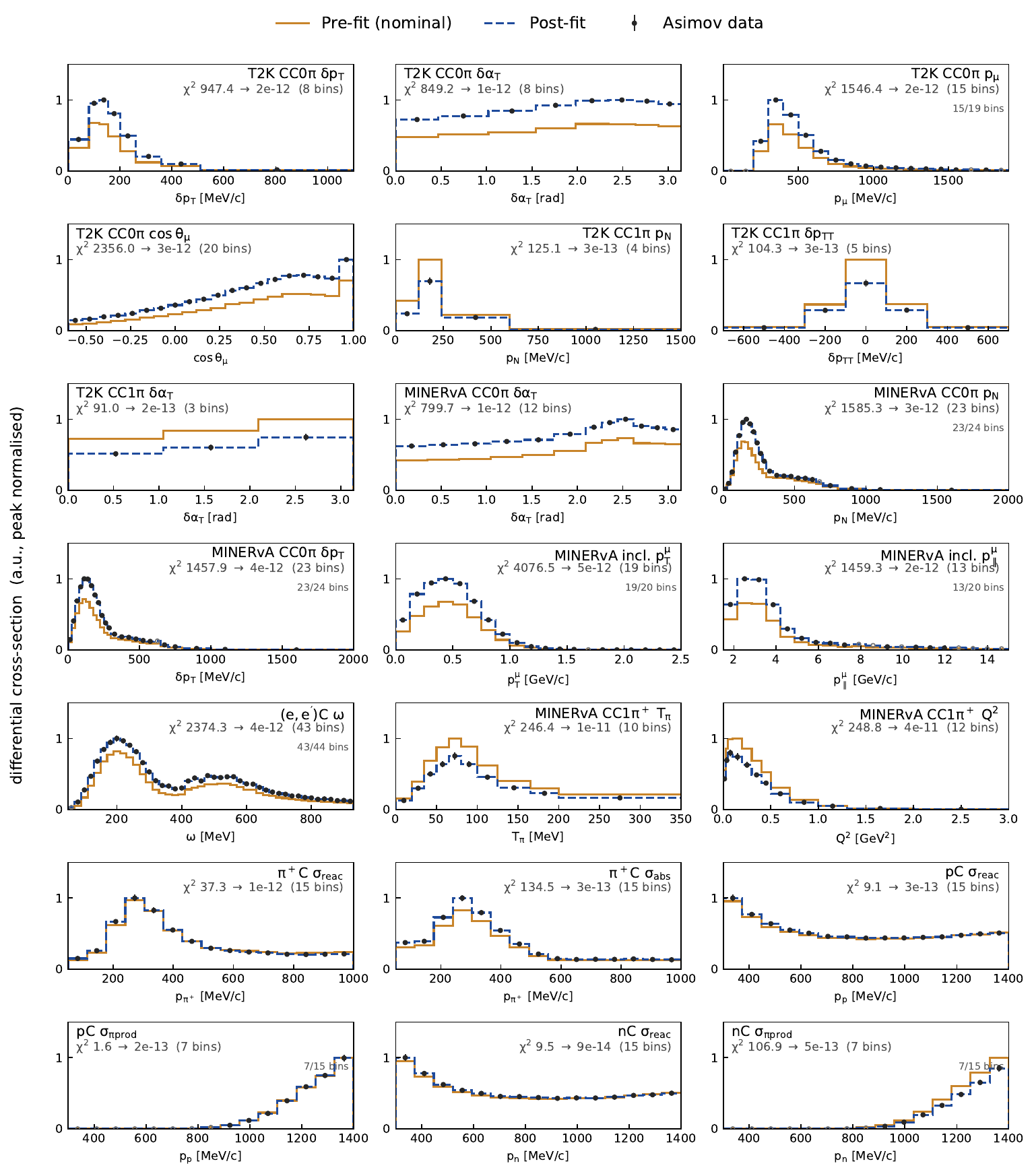}
  \caption{Every observable entering the fit, before and after it. Points are
    the Asimov data with the per-bin uncertainty used in the fit; hollow points
    are the bins removed by the Monte Carlo error mask. The pre-fit curve is
    the prediction at nominal and the post-fit curve the prediction at the
    recovered parameters. Each panel is normalised to its own peak and quotes
    its $\chi^2$ before and after the fit over the retained bins. For the two
    observables whose upper edge is not a kinematic limit, the overflow bin is
    omitted both from the drawing and from the $\chi^2$.}
  \label{fig:fit_rates}
\end{figure*}

This section illustrates how a differentiable event generator constructed as
described above can be used directly for parameter inference and uncertainty
quantification. In \adonis, the derivative of any observable built from the
events is available by construction (Sec.~\ref{sec:adonis_events}), so any
differentiable loss can be optimised directly without external response
functions or finite-difference derivatives. We demonstrate this in
cross-section space, where experiments routinely report their measurements,
and show that the same framework supports the standard inference machinery
used in HEP, from fitting and Gaussian uncertainties to profiling and
marginalisation. Our focus here is on demonstrating this machinery, thereby
enabling applications including tests of the self-consistency of published
measurements, global model tuning, and the extraction and Gaussianity testing
of correlated cross-section model constraints; pursuing these applications is
left to dedicated future studies.

We demonstrate these capabilities using the same $21$ observables and $17$
parameters identified in Sec.~\ref{sec:jacobian}. For the closure test, the
parameters are randomly displaced within their physical ranges to define an
injected truth, and the nominal event bank is reweighted to the corresponding
prediction. Each bin is assigned the same $5\%$ uncertainty used in
Sec.~\ref{sec:jacobian}. No informative parameter priors are imposed, although
all parameters remain restricted to their physical ranges. For simplicity, we
use a Gaussian $\chi^2$ throughout; the same differentiable prediction can
enter any differentiable likelihood.

Parameter optimisation is performed with a Gauss--Newton
method~\cite{Nocedal:2006} using the exact Jacobian supplied by \adonis.
Sec.~\ref{sec:fit_performance} describes the method in more detail and compares
its performance with MIGRAD~\cite{James:1975dr, iminuit}.

Figure~\ref{fig:fit_rates} shows the result in observable space. The injected
truth is sufficiently displaced from nominal that the pre-fit prediction
visibly differs from the data wherever the shifted parameters have an effect,
while the post-fit prediction recovers it. Because the closure dataset is
generated exactly at the injected truth, the residual at the minimum vanishes
by construction.

\begin{figure}[htbp]\centering
  \includegraphics[width=\linewidth]{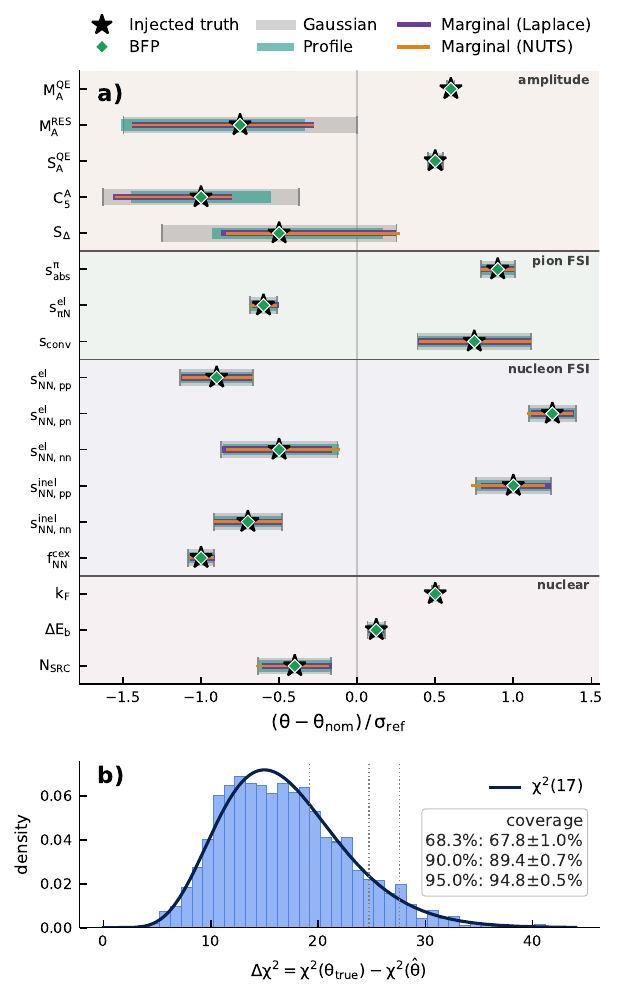}
  \caption{(a) Recovery of the $17$ fitted parameters, in units of the reference
  width $\delta\theta_k$ defined in Sec.~\ref{sec:jacobian}. Every parameter
  returns to its injected truth. The four overlaid bars show the Gaussian,
  profile, Laplace-marginal and directly sampled marginal intervals, all at
  $68\%$ and all obtained from the same dataset. The direct marginal is obtained
  with the No-U-Turn Sampler (NUTS). (b) Distribution over the toy ensemble of
  $\Delta\chi^2$ between the true and fitted parameters, compared with the
  $\chi^2$ distribution for $17$ degrees of freedom predicted by Wilks'
  theorem. The observed coverage is quoted at three
  confidence levels.}
  \label{fig:closure}
\end{figure}

The same differentiable prediction can also be used directly for uncertainty
quantification, without restricting the analysis to the local Gaussian
approximation around the fitted minimum. Panel~(a) of
Fig.~\ref{fig:closure} compares four $68\%$ intervals obtained from the same
dataset. The Gaussian interval uses the local covariance at the minimum; the
profile evaluates the likelihood away from it along the parameter of
interest; the Laplace marginal also accounts for the volume of the nuisance
directions; and direct sampling explores the full parameter space. These
provide increasingly non-local uses of the same underlying prediction and
allow their assumptions to be tested against one another.

\begin{figure*}[t]\centering
  \includegraphics[width=0.7\linewidth]{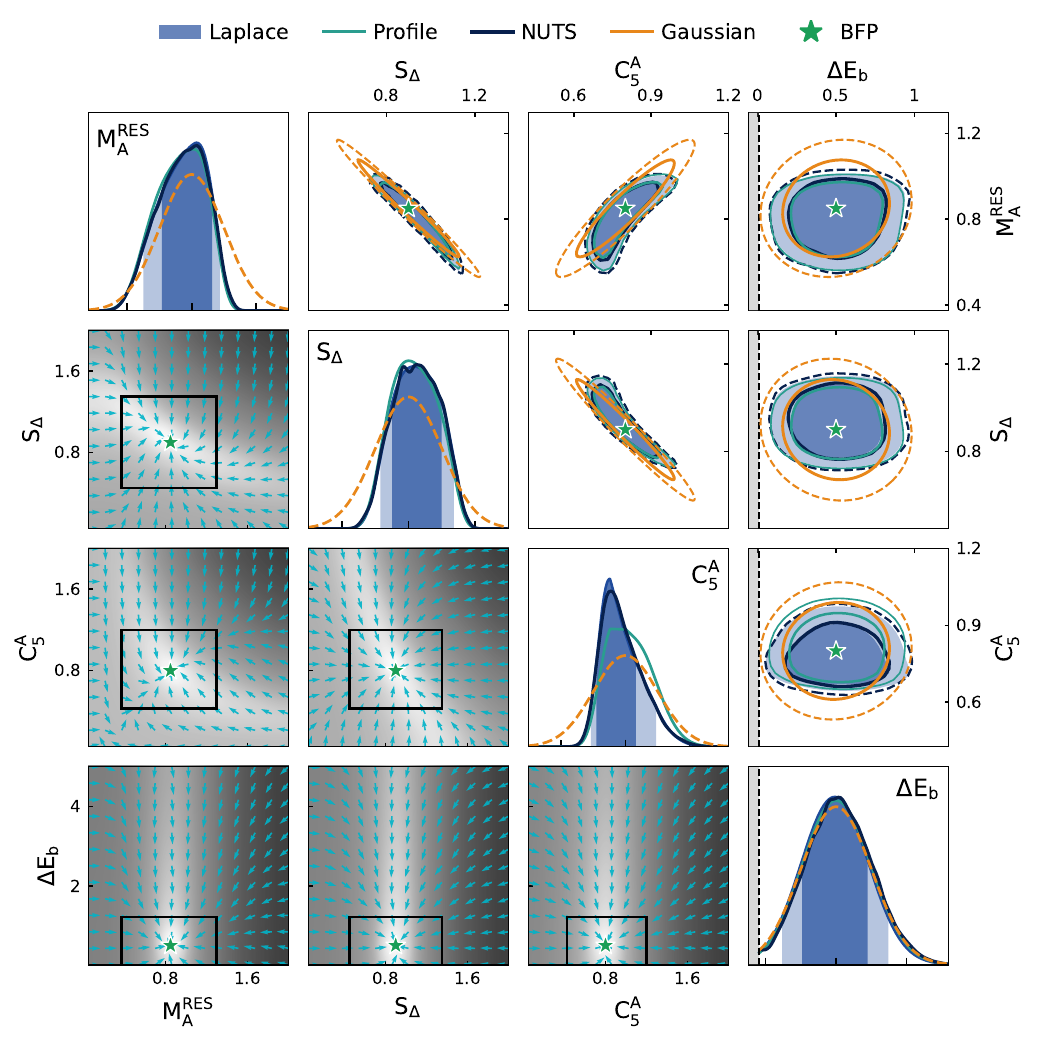}
  \caption{Four parameters of the fit in two dimensions. Below the diagonal,
    the parameter pairs are shown over a wide range of physical values, with
    arrows giving the local Gauss--Newton step direction over
    $\log_{10}\Delta\chi^2$ in greyscale. The rectangles mark the regions
    shown above the diagonal, where the $68\%$ and $90\%$ regions of four
    uncertainty constructions are compared. The diagonal panels show the
    corresponding one-dimensional distributions. The green star marks the
    best-fit point (BFP), and the grey band in $\Delta E_b$ denotes values outside
    the physical range.}
  \label{fig:corner}
\end{figure*}

To demonstrate that the differentiable treatment extends beyond the local
curvature, we construct a profile likelihood for each parameter by fixing
$\theta_k$ and re-minimising over the remaining $16$ parameters,
\begin{equation}
\label{eq:profile}
\chi^2_{\rm prof}(\theta_k)
=
\min_{\bm{\theta}_{\rm nuis}}
\chi^2(\theta_k,\bm{\theta}_{\rm nuis}).
\end{equation}
Each profile is evaluated at $13$ nodes, with its range extended independently
on either side until $\Delta\chi^2_{\rm prof}>25$ or a physical boundary is
reached. This adaptive range matters for shallow directions, where several
Gaussian widths need not exhaust the likelihood. To compare with the marginal
intervals on a common $68\%$ mass basis, we normalise
\begin{equation}
\mathcal{L}_{\rm prof}(\theta_k)
\propto
e^{-\Delta\chi^2_{\rm prof}(\theta_k)/2}
\end{equation}
over the scanned range and quote its shortest $68\%$ region.

The same framework also supports marginalisation, where nuisance parameters
are integrated rather than optimised. The marginal likelihood is
\begin{equation}
\label{eq:marginal}
\mathcal{L}_{\rm marg}(\theta_k)
=
\int
\mathcal{L}(\theta_k,\bm{\theta}_{\rm nuis})
\,d\bm{\theta}_{\rm nuis}.
\end{equation}
A second-order expansion about the conditional best fit gives the Laplace
approximation
\begin{equation}
\label{eq:laplace}
\mathcal{L}_{\rm marg}(\theta_k)
\propto
\mathcal{L}_{\rm prof}(\theta_k)
\sqrt{\det V_{\rm nuis}(\theta_k)},
\end{equation}
where $V_{\rm nuis}$ is the local nuisance-parameter covariance. The profile
measures the height of the likelihood ridge, while the determinant measures
its transverse volume; the two differ whenever that volume changes along the
ridge.

The second derivatives available from the same model allow this
marginalisation to use the exact local curvature rather than reusing the
Gauss--Newton, or Fisher, approximation of
Eq.~\ref{eq:adonis_fisher}. At each profile point we evaluate
$V_{\rm nuis}$ from the exact Hessian by nested automatic differentiation.
This matters away from the global minimum, where the residual is generally
non-zero and the residual-dependent contribution to the Hessian need not
vanish. Second-order information can therefore be used where the statistical
construction requires it, without building a separate numerical response
model.

The analysis is not restricted to marginalisation through a local expansion.
We also sample the same $17$-dimensional likelihood directly with the
No-U-Turn Sampler (NUTS), an MCMC algorithm based on Hamiltonian Monte
Carlo~\cite{Hoffman:2011ukg}, assuming a uniform density over the physical
parameter ranges of Sec.~\ref{sec:adonis_reweighting} and no additional prior.
The four constructions can then be compared directly: Gaussian versus profile
tests the local quadratic description along the parameter of interest;
profile versus Laplace tests the importance of nuisance volume; and Laplace
versus NUTS tests the quadratic approximation used to describe that volume.
All four are obtained from the same differentiable prediction. Gradient
information therefore does not confine the analysis to Gaussian or locally
approximated uncertainties, but remains available throughout profiling and
full marginalisation.

The same fitting machinery remains well behaved when applied repeatedly to
statistically fluctuated datasets. Panel~(b) of Fig.~\ref{fig:closure}
repeats the fit for $2000$ toys, obtained by fluctuating the same injected
prediction with the per-bin uncertainties, and forms
\begin{equation}
\Delta\chi^2
=
\chi^2(\bm{\theta}_{\rm true})
-
\chi^2(\hat{\bm{\theta}}).
\end{equation}
The ensemble follows the $\chi^2$ distribution with $17$ degrees of freedom
expected from Wilks' theorem~\cite{Wilks:1938dza}. The measured coverages are
$67.8\pm1.0\%$, $89.4\pm0.7\%$, and $94.9\pm0.5\%$, compared with nominal
$68.3\%$, $90\%$, and $95\%$, respectively, and the median $\chi^2$ per
degree of freedom is $0.992$. The differentiable treatment therefore
reproduces not only the injected minimum, but the expected statistical
behaviour of an ensemble of fits without outliers. Carrying exact derivatives
through the generator introduces no observable penalty in fit convergence or
frequentist coverage.

Physical boundaries are common in HEP fits and provide an important test of
the inference machinery. The injected value of $\Delta E_b$ lies close to the
physical wall where the spectral-function reweight becomes flat, producing a
truncated and strongly asymmetric likelihood. Such a boundary is precisely a
regime in which Gaussian approximations cease to apply; when the true value
lies on it, the asymptotic likelihood-ratio distribution is described by the
corresponding Chernoff mixture rather than the usual Wilks
result~\cite{Chernoff:1954}. No modification of the differentiable model is
required: the physical range is part of the minimisation domain, the profile
terminates at the boundary, and NUTS remains within it. The same gradient
information can therefore be used unchanged in fits involving physical
limits and strongly non-Gaussian uncertainties.

The derivative information is also usable across correlated multi-parameter
regions, not only in the immediate neighbourhood of the best fit.
Figure~\ref{fig:corner} shows this for the three parameters sharing the
resonant axial block and for $\Delta E_b$ with its physical boundary. Above
the diagonal, agreement of the four uncertainty constructions identifies
directions where the local Gaussian description is sufficient; their
separation localises the non-Gaussianity to particular parameter combinations
and, through the comparisons above, identifies its origin.

Below the diagonal, the same derivatives are evaluated over the full
physically allowed parameter space. The arrows show the local Gauss--Newton
step from each point: near the minimum they determine the covariance of the
inferred region, while farther away they determine the direction in which the
fit moves. Their smooth behaviour across the full range shows that the
gradient supplied by \adonis is available throughout the parameter space,
rather than being a response constructed only around the nominal point or the
best fit. The same differentiated model therefore describes both the
optimisation landscape and the uncertainty around its solution.

Taken together, these tests show that the gradient information supplied by
\adonis can be carried through the full inference workflow without narrowing
the statistical tools available to the analysis. At the analysis level, this
realises the same advantage that motivates built-in differentiability
throughout the generator: additional information and computational leverage
are gained without imposing a corresponding restriction on how the physics
inference is performed.

\subsection{Computational cost}
\label{sec:fit_performance}

Built-in derivative information can be exploited at every computational stage
of the inference demonstrated above. We benchmark its impact in three
representative tasks: locating the best fit, evaluating the curvature entering
uncertainty quantification, and sampling the full posterior with MCMC. Unless otherwise stated, all performance tests are carried out in double
precision on the same NVIDIA A100 GPU. JAX compilation is excluded from the reported timings; for MCMC, warm-up
and adaptation are also excluded from the sampling times. Optimisation times are
quoted as medians over repeated statistically fluctuated fits. In each case the
statistical target is unchanged; what changes is how efficiently the information
required by the calculation can be obtained.

For optimisation, we compare three choices on the same objective:
MIGRAD~\cite{James:1975dr, iminuit} without derivative information, MIGRAD
supplied with the exact gradient from \adonis, and Gauss--Newton (GN), which
uses the full Jacobian supplied by \adonis to construct a curvature-aware step
without requiring second derivatives of the generator. For GN, the step is
controlled by a trust region whose size is updated according to how accurately
the local model predicts the reduction actually achieved in $\chi^2$.
Physical parameter bounds are incorporated into the step itself, slowing
directions as they approach a boundary and reflecting otherwise infeasible
steps back into the allowed domain~\cite{Branch:1999trf,
Virtanen:2019joe}.

For the representative $17$-parameter fit with statistical fluctuations,
using $1.25\times10^{5}$ events per sample, GN converges in a median
$1.6$~s, compared with $4.8$~s for MIGRAD supplied with the exact
autodifferentiated gradient and $12.3$~s without one. Supplying derivative
information therefore accelerates the fit substantially.  In a separate hardware comparison, the same fit was approximately $120$ times slower on a single AMD EPYC CPU core than on the A100, excluding compilation
in both cases.

The number of events and the number of parameters $n$ affect the computational
cost differently. At fixed $n$, the arithmetic workload of each model or
derivative pass grows linearly with the number of events; the total cost then
depends on how many such passes the inference algorithm requires. Increasing
$n$ additionally changes the cost of obtaining the derivative information
required by the optimiser. MIGRAD without supplied derivatives must estimate
the gradient numerically, so a central finite-difference estimate requires
approximately $2n$ evaluations of the objective. When MIGRAD is supplied with
derivatives from \adonis, automatic differentiation instead obtains all
components of the gradient of the scalar objective together, avoiding a
separate pair of model evaluations for each parameter. The same advantage
applies to GN, although it uses the full residual Jacobian rather than only
$\nabla\chi^2$.

Repeating the statistically fluctuated fits while increasing the number of
parameters from $2$ to $17$ makes this distinction explicit. Over the range
tested, the measured GN wall time grows approximately as $n^{0.8}$,
gradient-supplied MIGRAD as $n^{0.9}$, and MIGRAD with numerical derivatives as
$n^{1.5}$. Both autodifferentiated approaches therefore scale much more mildly
than numerical differentiation as the parameter space grows.

The same advantage propagates to uncertainty quantification, where the
appropriate construction depends on the dimensionality of the reported result
and the available computational budget. The local Gaussian covariance is the
least expensive option and is already available from the Jacobian used by GN.
For low-dimensional non-Gaussian results, profile scans require a conditional
fit at every grid point, so each point benefits from the same derivative
information as the central fit.

Marginalisation on the same low-dimensional grids additionally requires the
nuisance curvature entering Eq.~\ref{eq:laplace}. Reusing the
Gauss--Newton/Fisher matrix of Eq.~\ref{eq:adonis_fisher} is essentially free,
but away from the minimum it omits the residual-dependent part of the exact
Hessian. At representative profile points in this fit, that approximation
changes $\log\det V_{\rm nuis}$ by up to about $0.25$, motivating the exact
calculation used above. Automatic differentiation assembles the exact Hessian
with $\mathcal{O}(n)$ Hessian--vector products, compared with
$\mathcal{O}(n^2)$ objective evaluations for a finite-difference calculation.
This requires $48$ event passes and about $0.18$~s per point, compared with
roughly $285$ passes and $1.35$~s for MINUIT HESSE, making the exact
autodifferentiated calculation about seven times faster.

For fully correlated high-dimensional inference, HEP analyses commonly turn
to Markov chain Monte Carlo (MCMC). Here the analogous distinction is between
random-walk algorithms such as Metropolis--Hastings (MH), which require only
likelihood evaluations, and Hamiltonian methods such as NUTS, which also exploit its gradient. Since \adonis supplies that gradient
directly, gradient-based MCMC can use it without constructing numerical
derivatives of the generator.

Because MCMC efficiency depends strongly on the posterior geometry, we compare
MH and NUTS on the same $17$-parameter posterior. Both sample the same
likelihood, physical bounds and dataset through a shared objective. MH uses a
Gaussian random-walk proposal preconditioned by the Laplace covariance, with
its scale adapted to a target acceptance of $0.234$~\cite{roberts1997weak};
NUTS targets $0.80$ and uses dual averaging with windowed metric adaptation as
implemented in Stan~\cite{hoffman2014no, carpenter2017stan}. Efficiency is compared only
after both samplers satisfy a rank-normalised, folded
split-$\hat{R}<1.01$ convergence criterion~\cite{vehtari2021rank} and their
one-dimensional marginals are found to agree.

Sampling efficiency is quantified through the effective sample size (ESS),
estimated with the initial monotone positive sequence~\cite{geyer1992practical}.
Because ESS is parameter dependent, we quote both its minimum across
parameters, which captures the slowest-mixing direction, and its median. Per
unit wall-clock time, NUTS yields $2.0$ times the minimum ESS of MH and
$1.4$ times the median ESS on this posterior. The gain rises to $3.4$ for the
tail ESS at the $5\%$ and $95\%$ quantiles bounding a $90\%$ credible
interval. Gradient-based NUTS therefore produces more statistically effective
posterior samples per unit time than the tuned random-walk MH comparison,
with the largest gain in the tails relevant to quoted interval endpoints.

\section{Unfolding}
\label{sec:unfolding}

The previous section used the differentiable prediction directly in truth space
to constrain interaction-model parameters. The same machinery can also be
carried through the detector response, allowing the signal in true kinematics
to be extracted from reconstructed data while fitting flux, detector and
interaction-model nuisance parameters simultaneously. This is the setting of a
cross-section measurement: experiments observe event distributions in
reconstructed kinematics, from which the corresponding signal rates in true
kinematics must be inferred after accounting for detector effects and
backgrounds, before normalisation by the integrated flux and number of targets.
Exploiting the differentiability of \adonis, we illustrate this extraction with
a template-fit unfolding, the standard approach used in T2K cross-section
measurements\footnote{A direct comparison with iterative Bayesian
unfolding~\cite{DAgostini:1994fjx}, the other widely used approach in the field,
is given in Ref.~\cite{T2K:2016jor}.}. 

We demonstrate this by mirroring the T2K $\nu_\mu$ CC0$\pi$
double-differential measurement of Ref.~\cite{T2K:2020sbd}, in muon momentum
and angle, $d^2\sigma/dp_\mu\,d\cos\theta_\mu$. We adopt the same signal
definition: a $\nu_\mu$ charged-current interaction with exactly one $\mu^-$,
no pions or other mesons after FSI, any number of nucleons, and no restriction
on the muon kinematics. We also retain the same $58$ truth bins, while
splitting each momentum interval in two at reconstruction level to obtain
$116$ reconstructed bins. A simple stochastic detector model applies a
$10\%$ momentum smearing and a $5^\circ$ angular smearing event by event. To
introduce a background component, charged pions below $400$~MeV/$c$ are reconstructed with $50\%$ efficiency,
implemented as an event-by-event random selection, so missed pions allow CC1$\pi$ events
to enter the CC0$\pi$ selection, giving a signal purity of $0.915$. The
prediction is normalised to an exposure corresponding to approximately
$170\,000$ selected events, while the underlying event bank provides about
$24$ times larger Monte Carlo statistics.

\begin{figure}[t]\centering
    \includegraphics[width=0.99\linewidth]{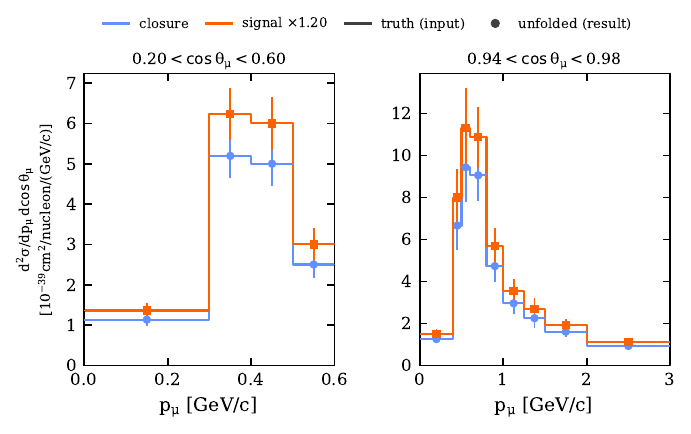}
    \caption{Unfolded T2K CC0$\pi$ double-differential cross section in two of
    the nine $\cos\theta_\mu$ slices, for a closure test and for a dataset in
    which the true signal is increased by $20\%$.}
    \label{fig:unfold_result}
\end{figure}

\begin{figure}[htbp]\centering
    \includegraphics[width=0.8\linewidth]{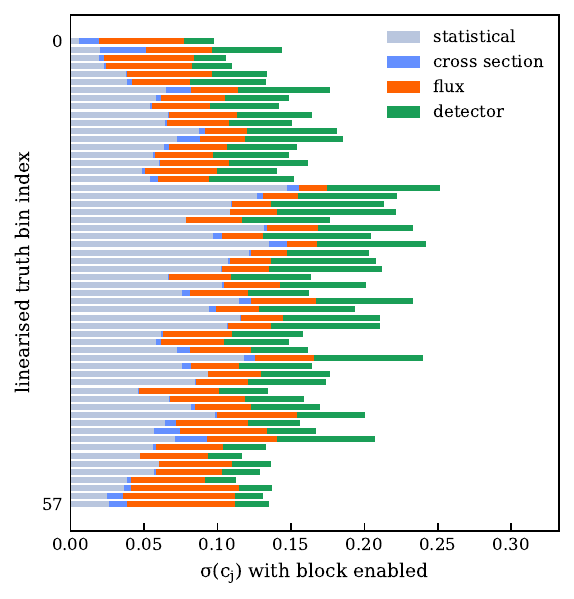}
    \caption{Uncertainty on the $58$ unfolded templates as statistical and
    systematic contributions are successively included.}
    \label{fig:unfold_budget}
\end{figure}

\begin{figure}[t]\centering
    \includegraphics[width=0.99\linewidth]{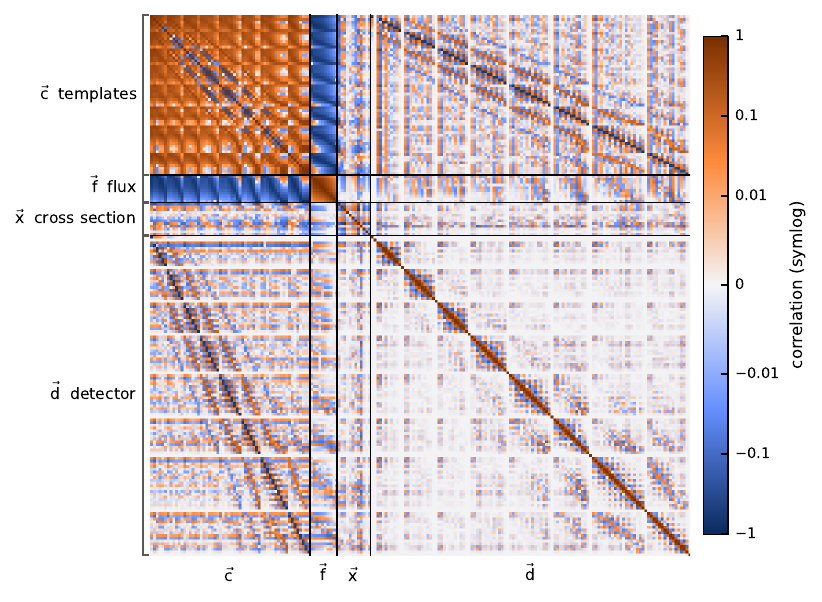}
    \caption{Post-fit correlation matrix of the $196$ fitted parameters:
    $58$ signal templates $\bm{c}$, $10$ flux parameters $\bm{f}$, $12$
    cross-section parameters $\bm{\theta}$ and $116$ detector parameters
    $\bm{d}$. The scale is symmetric-logarithmic, becoming linear for correlations below $0.01$ in
    magnitude.}
    \label{fig:unfold_corr}
\end{figure}

The prediction in reconstructed bin $i$ is
\begin{equation}
\label{eq:unfold_model}
\mu_i(\bm{c},\bm{f},\bm{\theta},\bm{d})
=
d_i\left[
\sum_{j,b} A_{ijb}\,c_j f_b
+
\sum_b B_{ib}(\bm{\theta})\,f_b
\right],
\end{equation}
where $A_{ijb}$ is the nominal signal template contributing to reconstructed
bin $i$ from truth bin $j$ and neutrino-energy bin $b$, while
$B_{ib}(\bm{\theta})$ is the corresponding background prediction. The $58$
template parameters $\bm{c}$ independently scale the signal truth bins and are
left unconstrained, so that the signal cross section is determined directly by
the fit rather than by the interaction model. The unfolded cross section in
each truth bin is therefore obtained by applying the fitted $c_j$ to the
corresponding nominal cross section.

The flux is described by $10$ parameters $\bm{f}$, each scaling all signal and
background events within an interval of true neutrino energy. The intervals are
$0$--$600$~MeV, four $100$~MeV bins between $600$ and $1000$~MeV, two
$200$~MeV bins between $1000$ and $1400$~MeV, two $300$~MeV bins between
$1400$ and $2000$~MeV, and an open bin above $2000$~MeV. For illustration,
they are assigned a correlated Gaussian prior with $10\%$ marginal
uncertainties and covariance
\begin{equation}
\Sigma_{bb'}=\sigma_b\sigma_{b'}
\exp\!\left(
-\frac{|\bar E_{\nu,b}-\bar E_{\nu,b'}|}{400~\mathrm{MeV}}
\right),
\end{equation}
where $\sigma_b=0.10$ and $\bar E_{\nu,b}$ is the centre of flux bin $b$.
For the open final bin, we set $\bar E_{\nu,b}=2150$~MeV by extrapolating the
width of the preceding finite bin, providing a finite representative energy
for the correlation kernel.

Of the $28$ cross-section parameters introduced in
Sec.~\ref{sec:adonis_reweighting}, we retain the $12$ whose variation on the
reference scales defined in Sec.~\ref{sec:jacobian} produces a non-negligible
change in the predicted background:
$k_F$, $S_\Delta$, $C_5^A$, $N_{\rm SF}$, $N_{\rm RES}$,
$M_A^{\rm RES}$, $f^{\rm cex}_{NN}$, $s^{\rm el}_{NN,pn}$,
$\Delta E_b$, $s_{\rm abs}^{\pi}$, $N_{\rm SRC}$ and
$s^{\rm el}_{\pi N}$. Concretely, we require
$\|(\partial B/\partial\theta_k)\,\sigma_k^{\rm ref}/\sqrt{\mu}\|_2\ge 1$
over the reconstructed bins, where $\mu$ is the nominal
signal-plus-background prediction and $\sigma_k^{\rm ref}$ the corresponding
reference variation. These parameters are also left unconstrained.

Finally, the $116$ detector parameters $\bm{d}$, one per reconstructed bin,
carry independent Gaussian $5\%$ normalisation uncertainties. The fit therefore
contains $196$ parameters in total and deliberately spans three different
nuisance treatments: unconstrained cross-section parameters, independently
constrained detector parameters, and correlated flux parameters.

An important feature of Eq.~\ref{eq:unfold_model} is that mapping the prediction
from true to reconstructed kinematics introduces no additional obstacle to
differentiability. For a given analysis definition, the outcomes of detector
smearing, event selection, and assignment to true and reconstructed bins are
fixed; they determine which events contribute to each bin, but not how their
weights respond to the fitted parameters. Importantly, these analysis choices
are themselves downstream of event generation. Different signal definitions,
selections and binnings can therefore be explored efficiently by reapplying
these downstream operations to the same event-level predictions and
derivatives.

Once this mapping is defined, the parameterisation used here is multilinear in
the template, flux, and detector parameters, $\bm{c}$, $\bm{f}$ and $\bm{d}$,
so their derivatives follow directly from the precomputed signal and
background tensors.

Figure~\ref{fig:unfold_result} shows the unfolded result for a closure dataset
and for a second dataset in which the true signal is increased uniformly by
$20\%$. In both cases the fit recovers the injected spectrum, demonstrating
that the unfolding pipeline behaves as expected.

The relative contributions to the uncertainty across the $58$ truth bins are
shown in Fig.~\ref{fig:unfold_budget}. Statistical, flux and detector
uncertainties all contribute appreciably, while the cross-section component
remains small because the high-purity sample limits the role of the interaction
model to the background prediction.

The corresponding post-fit correlations are shown in
Fig.~\ref{fig:unfold_corr}. The signal templates are anticorrelated with the
flux because both control the overall signal rate, while neighbouring templates
can be anticorrelated through detector migration between adjacent truth bins.
The detector block remains approximately diagonal, reflecting the independent
reconstructed-bin normalisations. Correlations involving the cross-section
parameters are more heterogeneous: different interaction parameters induce
distinct, generally non-linear distortions of the background and can therefore
correlate positively or negatively with different signal templates.

Taken together, these results show that the same differentiable framework can be used directly for unfolding.

\section{Conclusions}

We have presented \adonis, a fully differentiable neutrino interaction event
generator. By separating stochastic event generation from the differentiable
reweighting of the resulting event bank, \adonis propagates exact derivatives
through the nuclear ground state, hard-scattering amplitudes and intranuclear
cascade without modifying the underlying interaction physics. Its agreement
with ACHILLES across neutrino, electron and hadron probes demonstrates that
differentiability can be introduced without loss of physical fidelity.

Event reweighting and its gradients are available directly from the generator,
removing the need for external response functions to encode the dependence of
the prediction on the interaction-model parameters. We showed how the exact
Jacobian exposes which measurements constrain which parameters, where those
constraints originate in phase space, and how different probes break
degeneracies. The same differentiated prediction supports standard frequentist
and Bayesian parameter inference and uncertainty quantification, while
providing substantial computational gains for optimisation, curvature
evaluation and gradient-based MCMC. We further carried the same framework
through detector response in a $196$-parameter unfolding example, showing that
the gradient information remains directly usable in reconstructed-space
analyses.

\adonis is released as an open-source implementation of this
approach\footnote{\href{https://github.com/cesarjesusvalls/ADoNIS}
{https://github.com/cesarjesusvalls/ADoNIS}}, intended both as a practical tool
and as a reference for how differentiable neutrino event generators can be
constructed. The differentiable intranuclear cascade developed here is itself
modular and can be applied directly to the pre-FSI output of other event
generators, independently of the model used for the primary interaction.
Natural next steps for \adonis are the inclusion of two-body currents and
deep-inelastic scattering, using approaches analogous to those demonstrated in
this paper. Extending differentiability to hadronisation models such as
PYTHIA~\cite{Sjostrand:2006za, Sjostrand:2014zea} is also a valuable undertaking in its own right, given their
central role not only in DIS modelling for neutrino generators but throughout
high-energy physics. Importantly, these extensions are not prerequisites for
using the framework: until native differentiable implementations of two-body
currents and DIS are available, events from these channels can be taken from
existing generators, combined with the \adonis QE and RES samples, and
propagated through the \adonis cascade. The resulting prediction therefore
retains the full differentiable dependence of the \adonis QE and RES channels
on their model parameters, while preserving differentiability with respect to
the FSI parameters across all channels. Imported channels can additionally be
given differentiable parameter dependence through multiplicative event weights,
mirroring the flux and template parameters of Sec.~\ref{sec:unfolding}.

\section*{Acknowledgments}
The author acknowledges fruitful discussions with A. Lovato, O. Alterkait and
C. Wilkinson. This work was supported by compute credits from Anthropic's AI for Science program.

\bibliographystyle{apsrev4-1}
\bibliography{biblio}

\newpage

\appendix
\section{Additional validation distributions}
\label{app:validation}

\raggedbottom
\setlength{\intextsep}{8pt}

The remaining comparisons between \adonis and ACHILLES for the validation
distributions of Ref.~\cite{Isaacson:2025cnk} are shown below.

\begin{figure}[H]
  \centering
  \includegraphics[width=\linewidth]{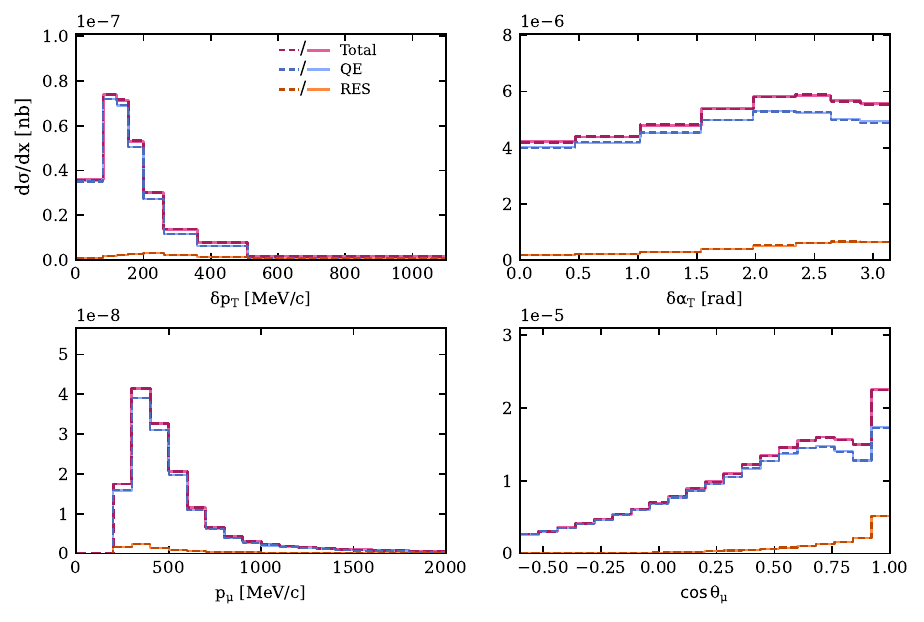}
  \caption{ADoNIS (solid) vs.\ ACHILLES (dashed). T2K CC0$\pi$-Np on $^{12}$C~\cite{T2K:2018rnz}: the TKI variables
  $\delta p_T$ and $\delta\alpha_T$, and the muon kinematics $p_\mu$ and
  $\cos\theta_\mu$.}
  \label{fig:val_t2k_cc0pi}
\end{figure}

\begin{figure}[H]
  \centering
  \includegraphics[width=\linewidth]{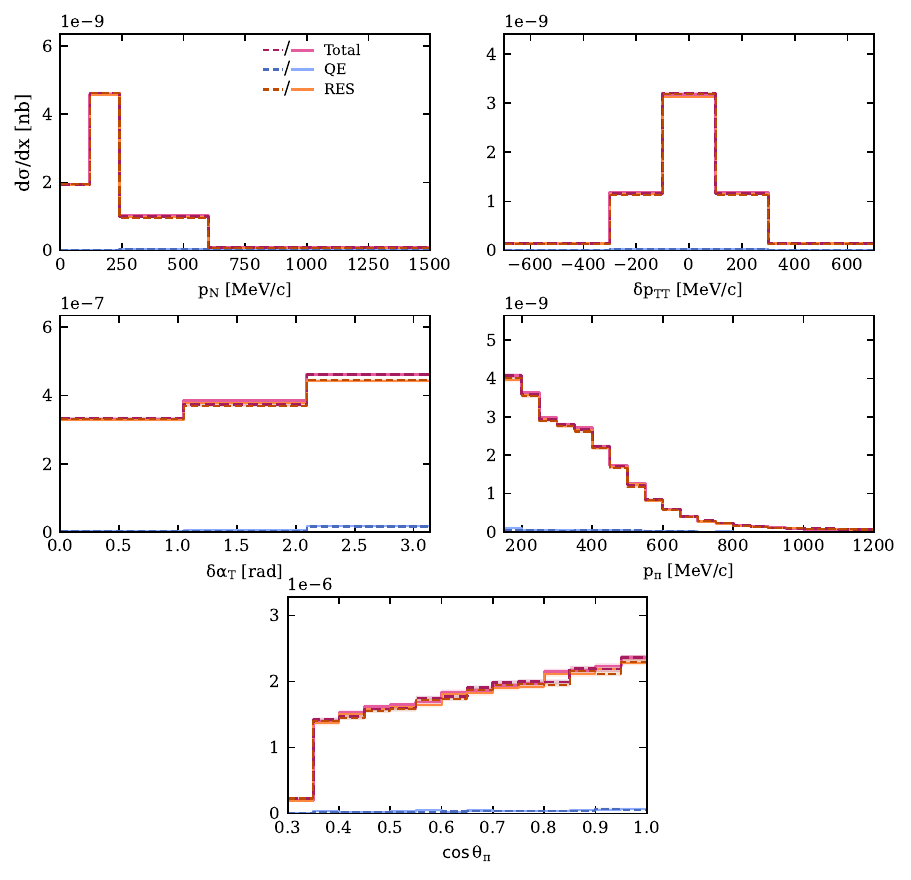}
    \caption{ADoNIS (solid) vs.\ ACHILLES (dashed). T2K CC1$\pi^+$Np on
    hydrocarbon~\cite{T2K:2021naz}: $p_N$, $\delta p_{TT}$,
    $\delta\alpha_T$, and the pion kinematics $p_\pi$ and
    $\cos\theta_\pi$.}
  \label{fig:val_t2k_cc1pi}
\end{figure}

\begin{figure}[H]
  \centering
  \includegraphics[width=\linewidth]{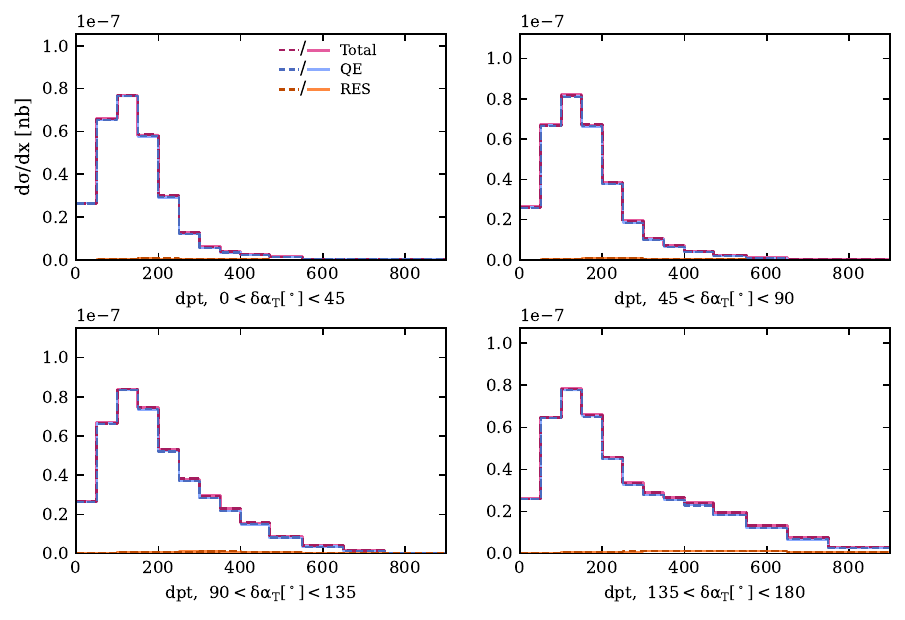}
  \caption{ADoNIS (solid) vs.\ ACHILLES (dashed). MicroBooNE CC1p0$\pi$ on $^{40}$Ar~\cite{MicroBooNE:2023tzj}:
  $\delta p_T$ in slices of $\delta\alpha_T$.}
  \label{fig:val_uboone_cc1p}
\end{figure}

\begin{figure}[H]
  \centering
  \includegraphics[width=\linewidth]{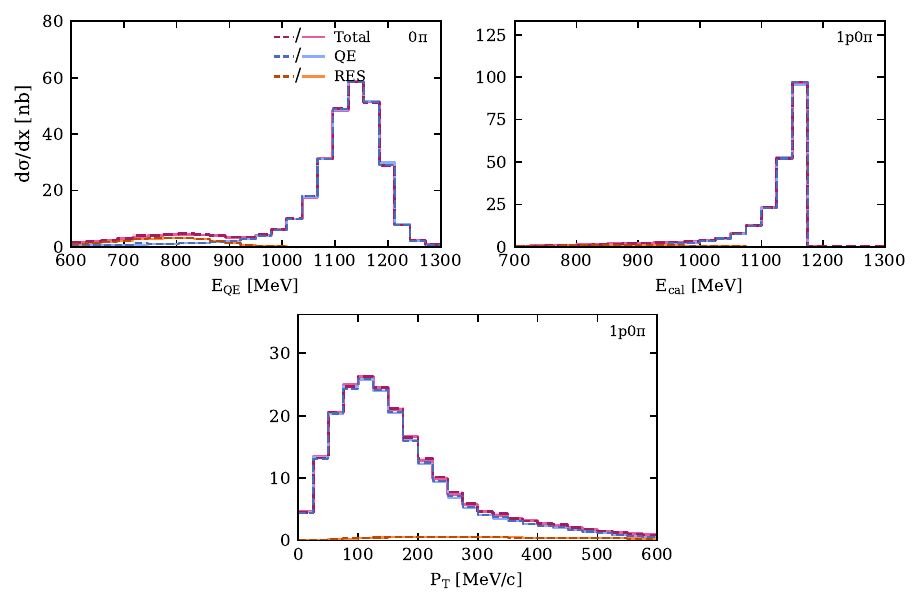}
  \caption{ADoNIS (solid) vs.\ ACHILLES (dashed). e4$\nu$ observables for $(e,e')$ on $^{12}$C at
  $1.159$~GeV~\cite{CLAS:2021neh}: the quasi-elastic and calorimetric
  neutrino-energy estimators and the transverse momentum imbalance.}
  \label{fig:val_e4nu}
\end{figure}

\clearpage
\end{document}